\documentclass[12pt,preprint]{aastex}

\usepackage{psfig,natbib}

\def\feka{Fe K$\alpha$}
\def\chandra{{\it Chandra}} 
\def\xmm{{\it XMM-Newton}} 
\def\asca{{\it ASCA}} 
 
\def\rxte{{\it RXTE}} 
\def\sax{{\it BeppoSAX}} 
 
\def\rosat{{\it ROSAT}} 
\def\ein{{\it Einstein}}

\def\ltsima{$\; \buildrel < \over \sim \;$}
\def\simlt{\lower.5ex\hbox{\ltsima}} 
\def\gtsima{$\; \buildrel > \over \sim \;$}
\def\simgt{\lower.5ex\hbox{\gtsima}} 

\begin{document}

\title{Long-Term, Continuous Monitoring of the Broad Line Radio Galaxies
3C~390.3 and 3C~120 With the Rossi X-Ray Timing Explorer}

\author{Mario Gliozzi\altaffilmark{\;1}, Rita
M. Sambruna\altaffilmark{\;1}, and Michael Eracleous\altaffilmark{\;2}}

\altaffiltext{1}{George Mason University, Dept. Of Physics \&
Astronomy, 4400 University Drive, Ms 3f3, Fairfax, Va 22030 (e-mail:
mario,rms@physics.gmu.edu).}

\altaffiltext{2}{The Pennsylvania State University, Department of
Astronomy And Astrophysics, 525 Davey Lab, State College, Pa 16802
(e-mail: mce@astro.psu.edu).}

\clearpage
\begin{abstract}
We present a study of the flux and spectral variability of the two
broad-line radio galaxies (BLRGs) 3C~390.3 and 3C~120, observed almost
daily with the {\it Rossi X-Ray Timing Explorer} (\rxte) for nearly
two months each in 1996 and 1997, respectively.  Our original
motivation for this study was to search for systematic differences
between BLRGs and their radio-quiet counterparts, the Seyfert
galaxies, whose temporal and spectral behavior is better studied. We
find that both 3C~390.3 and 3C~120 are highly variable, but in a
different way, and quantify this difference by means of a structure
function analysis. 3C~390.3 is significantly more variable than
3C~120, despite its jet larger inclination angle, implying either that the
X-ray variability is not dominated by the jet or that two different
variability processes are simultaneously at work in 3C~390.3. We
performed an energy-selected and time-resolved analysis based on the
fractional variability amplitude and found that the variability
amplitude of both objects is strongly anticorrelated with the
energy. This last result, along with the correlated change of the
photon index with the X-ray continuum flux, can be qualitatively
explained within the scenario of thermal Comptonization, generally
invoked for radio-quiet active galaxies. Moreover, the time-resolved
and energy-selected fractional variability analyses show a trend
opposite to that observed in jet-dominated AGN (blazars), suggesting
only a minor contribution of the jet to the X-ray properties of BLRGs.
Time-averaged spectral analysis indicates the presence of a strong, 
resolved \feka\ line with centroid at 6.4 keV 
and a weak ($\Omega/2\pi\simeq 0.1 - 0.4$) reflection
component in both objects. The overall PCA+HEXTE spectra are best
fitted with the constant density ionization model of Ross \& Fabian,
but with a modest ionization parameter. We performed a time-resolved
spectral analysis of 3C~390.3 with the aim of constraining the delay 
between \feka\ line and continuum variability; however, 
the limited signal-to-noise ratio of the 
line flux hampers a thorough study of the line variability. 
\end{abstract}

\section{Introduction}

Active Galactic Nuclei (AGN) are variable in every observable
waveband.  The X-ray flux exhibits variability on timescales shorter
than in any other energy band, indicating that the emission occurs in
the innermost regions of the central engine.  A fundamental open
question in the study of AGN concerns the difference between
radio-loud and radio-quiet objects. If, as widely believed, this
difference is not related to the nature of the host galaxy, but to
some intrinsic difference in the central engine (either connected to
the properties of the supermassive black hole or to the nature of the
accretion flow), then X-ray temporal and spectral variability can
provide constraints on the physical conditions in the inner parts of
the accretion flow and shed light on the radio-loudness dichotomy.

We have been carrying out a systematic study of radio-loud AGN at
X-ray wavelengths (Sambruna, Eracleous, \& Mushotzky 1999; Eracleous
Sambruna, \& Mushotzky 2000; Hasenkopf, Sambruna, \& Eracleous 2002)
to investigate in detail the properties of their nuclear emission and
the structure of their accretion flows.  While the X-ray timing
properties of Seyferts are relatively well known (e.g., Nandra et
al. 1997; Leighly 1999; Turner et al. 1999), in their radio-loud
counterparts, the Broad-Line Radio Galaxies (BLRGs), the variability
of the X-ray emission is not as well studied.  We targeted two
well-known BLRGs, 3C~390.3 and 3C~120, for intensive X-ray monitoring
because they are among the brightest BLRGs and are known to be
variable, based on previous studies with \ein\ and \rosat\ (Halpern
1985, Leighly \& O'Brien 1997).  We used \rxte\ to perform a long-term
monitoring of their flux and spectral variability, since it has a
unique combination of flexible scheduling, relaxed pointing
constraints, high throughput, and rapid slew speed.

Previous X-ray studies with {\it Ginga, ASCA, RXTE}, and \sax\
(Wo\'zniak et al. 1998, Sambruna et al. 1999, Eracleous et al. 2000,
Zdziarski \& Grandi 2001) show that BLRGs usually have weaker \feka\
emission lines and Compton ``reflection'' components than Seyfert
1s. While all the authors basically agree on these observational
results, their interpretation is still a matter of debate. In
particular, Wo\'zniak et al. (1998) suggested that the reprocessing
medium is located at a very large distance from the central engine and
identified the obscuring torus invoked in AGN unification schemes as a
plausible site. On the other hand, Eracleous et al. (2000) suggested
that the reprocessing medium is the outer part of the accretion disk,
which is much closer to the central engine than in the above picture
(at distances of order $10^3$ gravitational radii). These authors also
proposed that the central engines of BLRGs are fueled at a rate much
lower than the Eddington rate, with the result that their inner
accretion disks have the form of a vertically extended
advection-dominated accretion flow (ADAF; see, for example, Narayan,
Mahadevan, \& Quataert 1998 for a review) that illuminates the outer,
thin disk. The most recently suggested interpretation is by Ballantyne
et al. (2002), who noted that the weakness of the X-ray reprocessing
features is consistent with Compton ``reflection'' and fluorescence in
an {\it ionized} accretion disk. Therefore, they proposed that the
accretion disks of BLRGs are highly ionized as a consequence of a very
high accretion rate, which reaches a substantial fraction of the
Eddington rate. To illustrate the point these authors showed that the
\asca\ spectrum of 3C~120 can be fitted by a reprocessing model
employing an ionized accretion disk.

Distinguishing among the above interpretations is important since it
would help us understand the cause of the difference between
radio-loud and radio-quiet AGNs. With this in mind, we investigate
possible correlated variations of the Fe line and continuum flux in an
attempt to constrain the location of the reprocessor, and comment on
future observational tests.

The outline of the paper is as follows. In $\S~2$ we describe the
observations and data reduction. In $\S~3$ we study the temporal and
spectral variability of 3C~390.3 and 3C~120 using different
techniques: the structure function to determine the characteristic
time-scales of the sources; the fractional variability parameter to
quantify the timing behavior in energy-selected light curves and in
time-selected intervals; the hardness ratios to investigate the
spectral variability.  In $\S~4$ we analyze the spectral properties of
of the two objects by fitting different spectral models to the
time-averaged spectra and we also investigate the spectral evolution
of 3C~390.3 using time-resolved spectra. In $\S~5$ we discuss the
implications of the results of the temporal and spectral analysis.
Finally, in $\S~6$ we draw the main conclusions.
Throughout the paper we use a Friedman
cosmology with $H_0=75~{\rm km~s^{-1}~Mpc^{-1}~and}~q_0=0.5$ for the
computation of K-corrected luminosities.

\section{\rxte\ Observations And Data Reduction}

In Table~1 we list the basic properties of 3C~390.3 and 3C~120, namely
the redshift, the inferred inclination of the radio jet (see Eracleous
\& Halpern 1998 and references therein) and the column density of the
Galactic interstellar medium, along with the dates and times of the
observations.  3C~390.3 was monitored from 1996 May 17 to July 12,
with the following sampling: for the first 24 days the source was
regularly observed for roughly one hour once per day, for the
subsequent 10 days the sampling density was increased to two short
observations per day, then the source was again observed once per day
for 14 days, finally, during the last 10 days the observations were
carried out once every two days. 3C~120 was observed from 1997 January
10 to March 8, with a sampling pattern as even as possible
(i.e. approximately one observation per day).
The observations were carried out with the Proportional Counter 
Array (PCA; Jahoda et al. 1996), and the High-Energy X-Ray Timing 
Experiment (HEXTE; Rotschild et al. 1998) on \rxte.

The PCA data were screened according to the following acceptance
criteria: the satellite was out of the South Atlantic Anomaly (SAA)
for at least 30 min, the Earth elevation angle was $\geq 10^{\circ}$,
the offset from the nominal optical position was $\leq
0^{\circ}\!\!.02$, and the parameter ELECTRON-0 was $\leq 0.1$. The
last criterion removes data with high particle background rates in the
Proportional Counter Units (PCUs). The PCA background spectrum and
light curve were determined using the ${\rm L}7-240$ model developed
at the \rxte\ Guest Observer Facility (GOF) and implemented by the
program {\tt pcabackest} v.2.1b. This model is appropriate for
``faint" sources, i.e., those producing count rates less than 40 ${\rm
s^{-1}~PCU^{-1}}$. All the above tasks were carried out using the {\tt
FTOOLS} v.5.1 software package and with the help of the {\tt rex}
script provided by the \rxte\ GOF, which also produces response
matrices and effective area curves appropriate for the time of the
observation.  Data were initially extracted with 16 s time resolution
and subsequently rebinned at different bin widths depending on the
purpose for which they were employed.  The current temporal analysis
is restricted to PCA, STANDARD-2 mode, 2--11 keV, Layer 1 data,
because that is where the PCA is best calibrated and most
sensitive. Since PCUs 3 and 4 were occasionally turned off, only data
from the other three PCUs (0, 1, and 2) were used. All quoted count
rates are normalized to 1 PCU.  For the spectral analysis, in order to
increase the signal-to-noise ratio, we used data from all the
available PCUs (the spectra were combined with the help of the
software tool {\tt addspec}).

The HEXTE data, used in combination with the PCA data for the spectral
analysis, were also screened to exclude events recorded where the
pointing offset was greater than $0.02^{\rm o}$, and when the Earth
elevation angle was less than $10^{\rm o}$. The background in the two
HEXTE clusters is measured during each observation by rocking the
instrument slowly on and off source. Therefore, source and background
photons are included in the same data set and are separated into
source and background spectra according to the time they were
recorded. Due to their low signal-to-noise ratio, HEXTE data are used
only in the time-averaged spectral analysis, for the purpose of
constraining the reflection component of the broad band spectrum.
Average count rates of the PCA and HEXTE cluster 0, as well as
exposure times, are listed in Table~2.

\section{X-Ray Temporal Analysis}

Figure~\ref{figure:lc1B} shows the background-subtracted light curves
of 3C~390.3 and 3C~120 in the 2--11 keV energy range.  The observed
(2.1~${\rm s^{-1}~PCU^{-1}}$) mean PCA count rate for 3C~390.3 corresponds to a
2--10 keV flux of $2.6\times10^{-11}~{\rm erg~s^{-1}~cm^{-2}}$ and a
luminosity $L_{2-10}=1.6\times10^{44}~{\rm erg~s^{-1}}$.
For 3C~120 (5.2~${\rm s^{-1}~PCU^{-1}}$),
$F_{2-10}=6.1\times10^{-11}~{\rm erg~s^{-1}~cm^{-2}}$ and
$L_{2-10}=1.3\times10^{44}~{\rm erg~s^{-1}}$.
These luminosities were calculated assuming a
power-law spectral model with the best fit parameters that we 
present later on in this paper
(see Table 4).
3C~390.3 exhibits a trend of decreasing flux by a factor 2 in 20 days, with
smaller changes (a few percent) on timescales of 4 days or less,
superposed. 3C~120 shows erratic flux variations similar to
``flickering'', with small-amplitude intra-day ``flares'' superposed
on a roughly constant baseline. The most prominent feature in the
latter light curve is the presence of a sharp dip shortly
before the end of the monitoring campaign, with a count rate decrease
by a factor of 50\% in few days and a subsequence recovery to the
previous value in one day. As a check, we analyzed separately the 
light curves of the source and the background for each individual PCU
and concluded that the dip cannot be attributed to instrumental effects.
Moreover, the presence of remarkable short-time variability in  3C~120
is not surprising: \rosat\ observations (Grandi et al. 1997) have already
shown the presence of variations of the 0.1-2 keV flux by factors larger than
1 in $\sim 8000$ s, indicating that the region producing the photons 
cannot be larger than $10^{15}$ cm, 
provided that the X-ray emission is isotropic.
Throughout the temporal analysis section we assume for statistical
tests a level of significance of 90\%, meaning that there is a chance 
probability of 10\% that an hypothesis is erroneously rejected.

\subsection{Structure Function}

In order to quantify the different timing behavior of the two BLRGs,
we first carried out a structure function analysis (e.g. Simonetti et
al. 1985, Hughes et al. 1992). This is a method frequently used in
astronomy to quantify time variability without the problems
encountered in the traditional Fourier analysis technique in case of
unevenly sampled data. The first-order structure function is defined
as the mean deviation for data points separated by a time lag $\tau$:
\begin{equation} 
SF(\tau)=\langle [F(t)-F(t+\tau)]^2\rangle
\end{equation} 

One of the most useful features of the structure function is its
ability to discern the range of timescales that contribute to the
variations in the data set: the characteristic timescales of the
variability are identified by the maxima and slope changes in the
$\tau-SF$ plane. For a stationary random process the structure
function reaches a plateau state for lags longer than the longest
correlation timescale. In other words, the time at which the
structure function reaches a plateau state is interpreted as the
characteristic timescale of the source.

\begin{figure*}
\psfig{figure=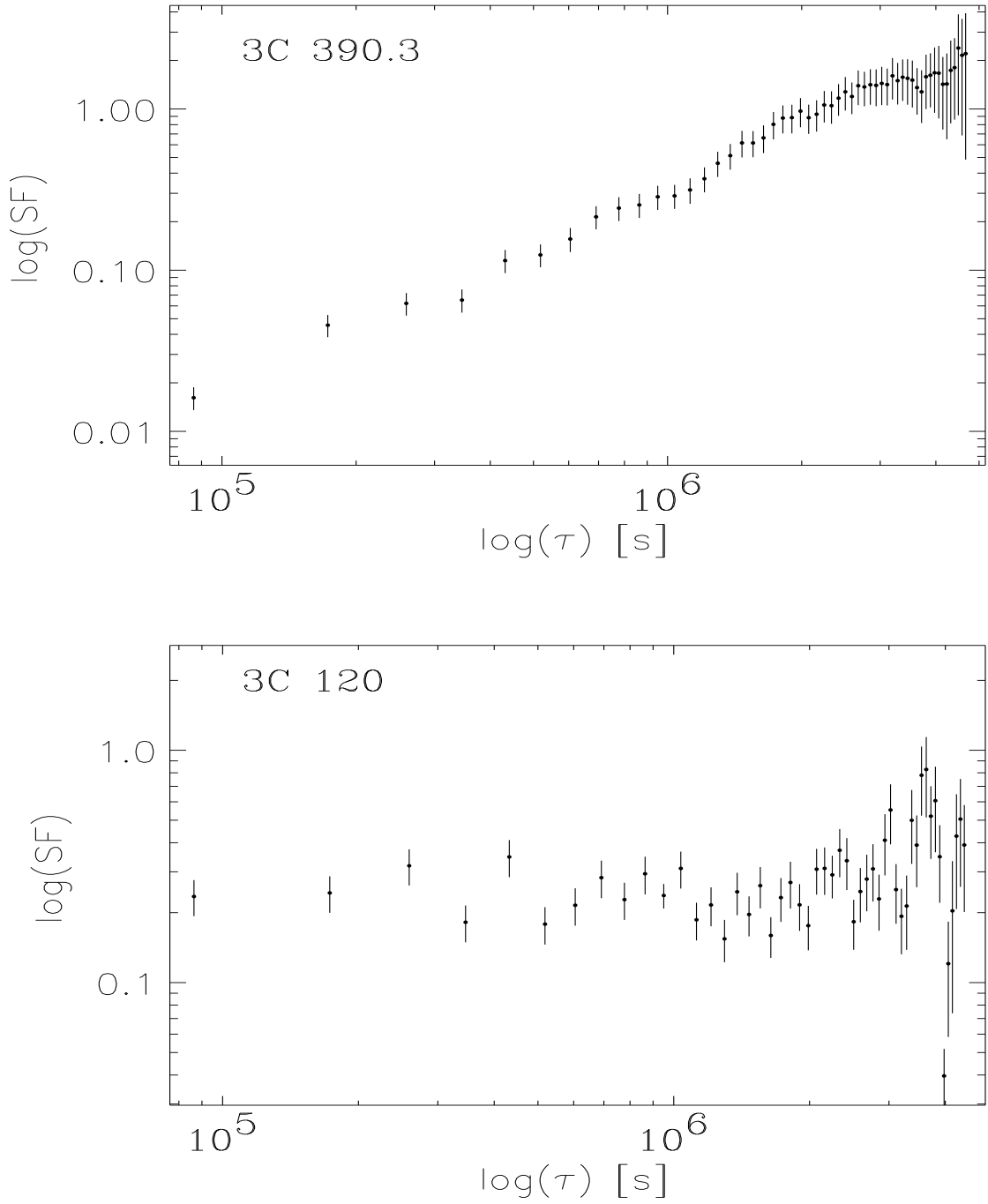,height=19cm,width=15cm,%
bbllx=60pt,bblly=94pt,bburx=490pt,bbury=766pt,angle=0,clip=}
\caption{Structure functions of 3C~390.3 (top panel) and 3C~120 (bottom panel).
\label{figure:sfB}}
\end{figure*}

Figure~\ref{figure:sfB} shows the structure functions of 3C~390.3 (top)
and 3C~120 (bottom panel). While the structure function of 3C~390.3
resembles the behavior expected from an ideal stationary random
process with a smooth increase until a plateau state is reached at
$\sim (2-2.5)\times10^6$~s ($\simeq$ 25-30 days), the structure
function of 3C~120 does not show any clear trend.  This reflects the
fact that 3C~390.3 showed a roughly smooth decay with a decaying
timescale of the order of 25 days, while 3C~120 did not show any long
term trend during the observational campaign.

\subsection{Fractional Variability}

A different way to quantify the variability properties of 3C~390.3 and
3C~120, without considering the time ordering of the values in the
light curves, is to calculate the fractional variability parameter
$F_{\rm var}$ (e.g. Rodriguez-Pascual et al. 1997, Edelson et
al. 2002). This is a common measure of the intrinsic variability
amplitude relative to the mean count rate, corrected for the effect of
random errors, i.e.,
\begin{equation}
F_{\rm var}={(\sigma^2-\Delta^2)^{1/2}\over\langle r\rangle}
\end{equation}
\noindent where $\sigma^2$ is the variance, $\langle r\rangle$ the
unweighted mean count rate, and $\Delta^2$ the mean square value of
the uncertainties associated with each individual count rate.

We first computed $F_{\rm var}$ for the whole light curves in the
2--11~keV energy band using 5760~s ($\sim$ one orbit) time bins, and
found 26.6\% and 8.5\% for 3C~390.3 and 3C~120, respectively. This
simply indicates that the long-term flux variability in 3C~390.3 is
larger than in 3C~120, as expected from the visual inspection of the
two light curves.

\begin{figure*}
\psfig{figure=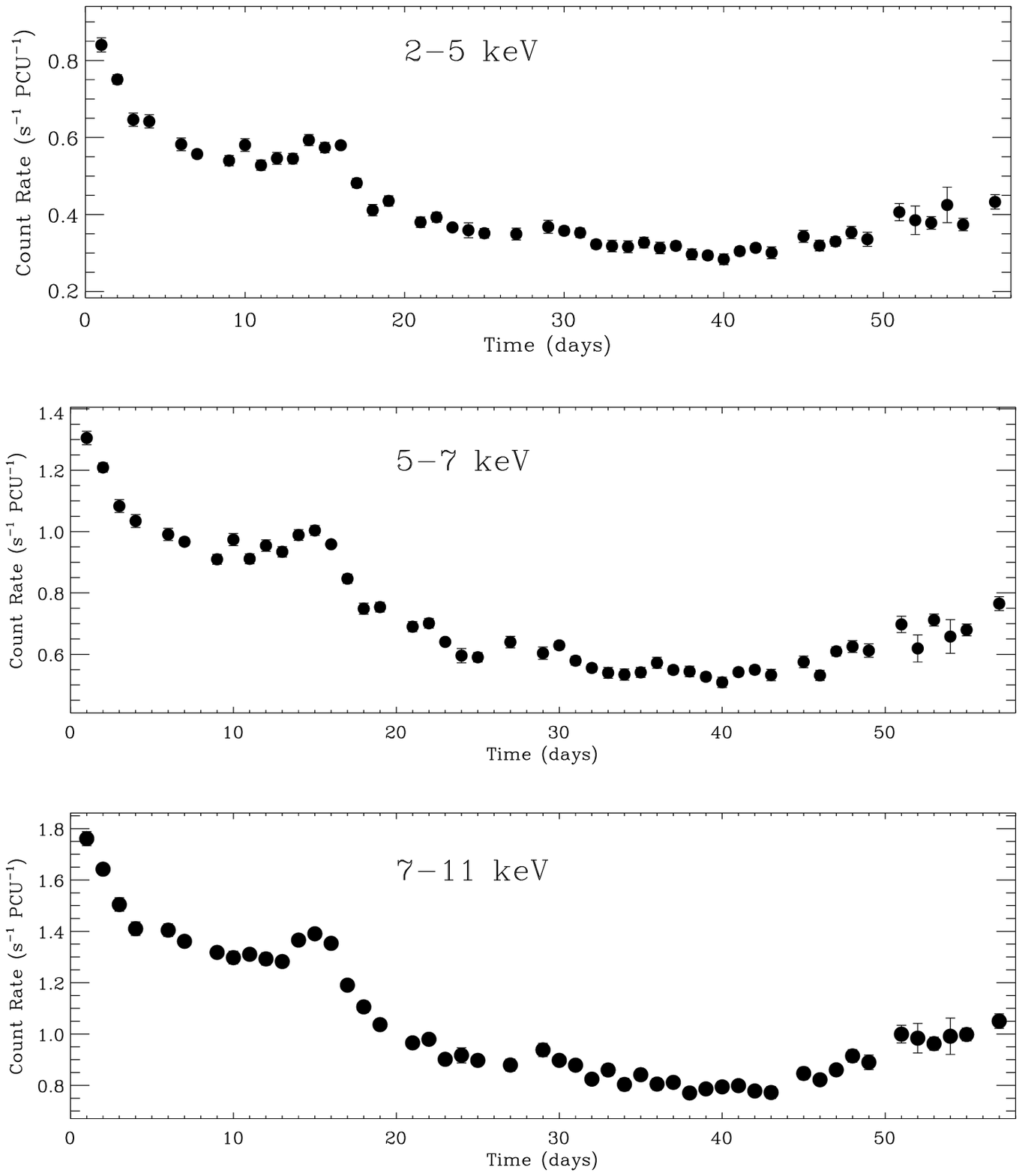,height=16.cm,width=14cm,%
bbllx=30pt,bblly=60pt,bburx=460pt,bbury=565pt,angle=0,clip=}
\psfig{figure=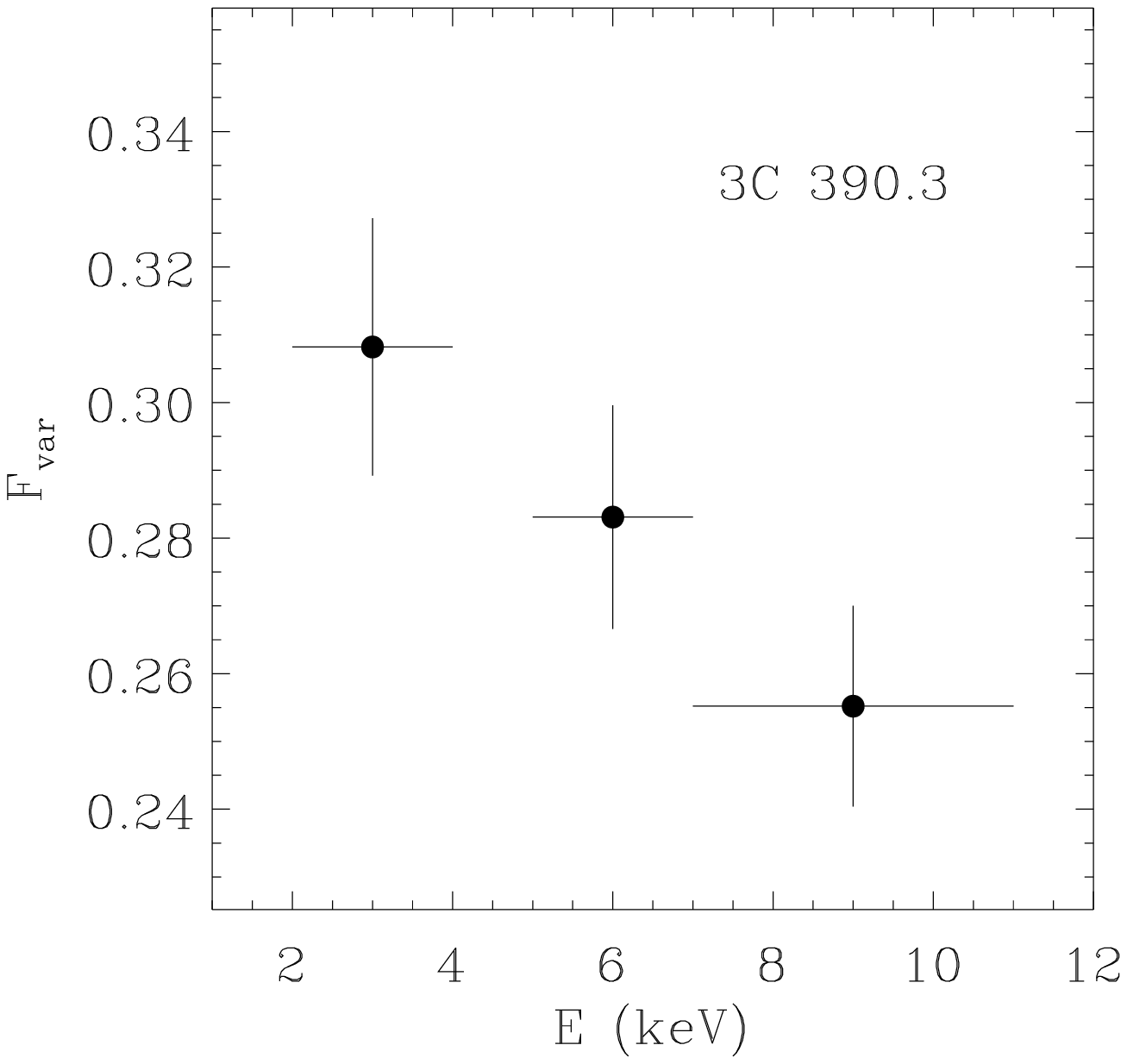,height=6.cm,width=10.5cm,%
bbllx=-140pt,bblly=69pt,bburx=420pt,bbury=430pt,angle=0,clip=}
\caption{X-ray light curves of 3C~390.3 from \rxte\ PCA observations
in (top to bottom) the 2-5 keV, 5-7 keV, 7-11 keV bands.  Time bins
are 5760 s ($\sim$ one \rxte\ orbit). The bottom plot shows the
fractional variability parameter versus the energy.
\label{figure:fvar390B}}
\end{figure*}

\begin{figure*}
\psfig{figure=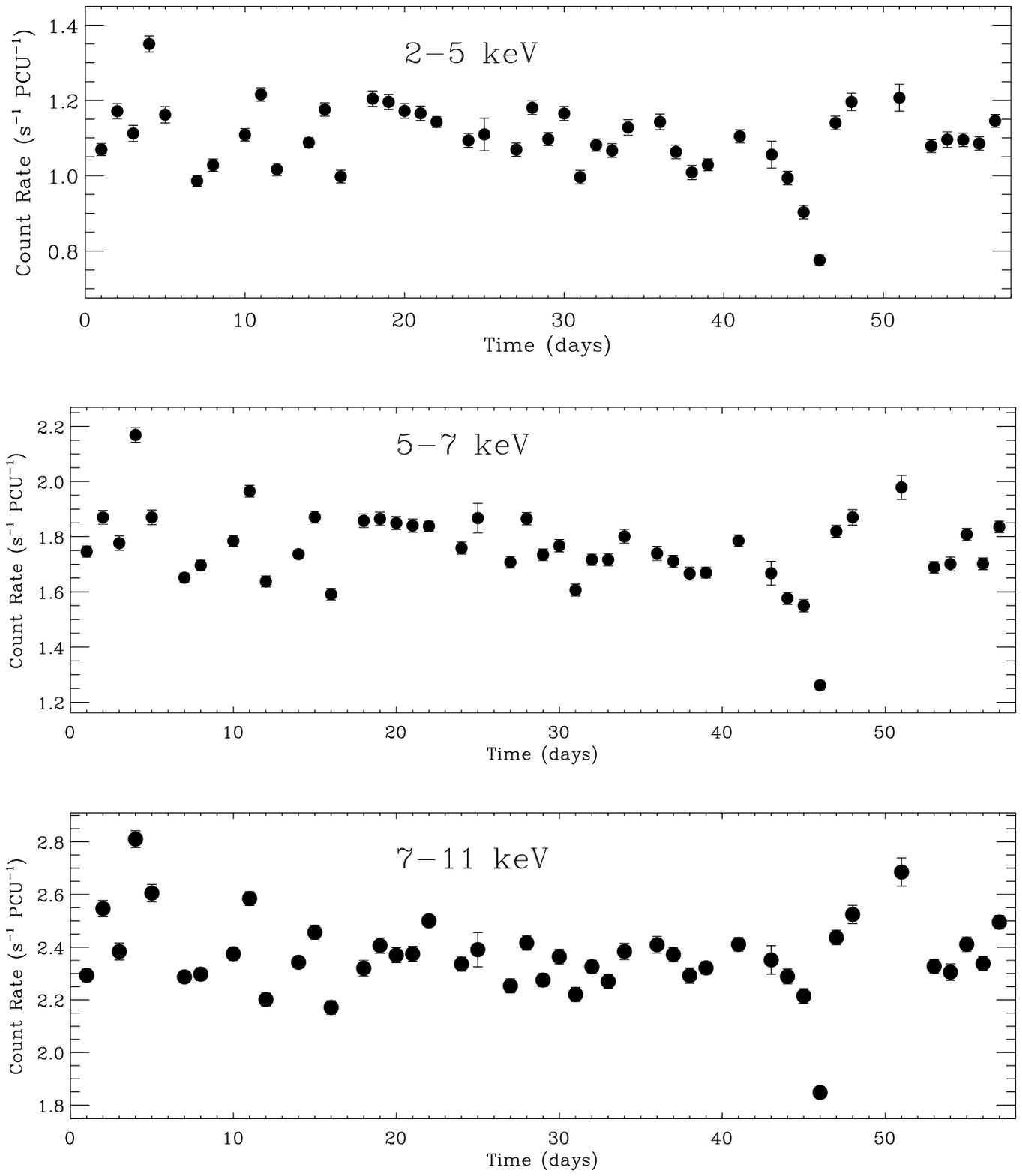,height=16.cm,width=14cm,%
bbllx=30pt,bblly=60pt,bburx=460pt,bbury=565pt,angle=0,clip=}
\psfig{figure=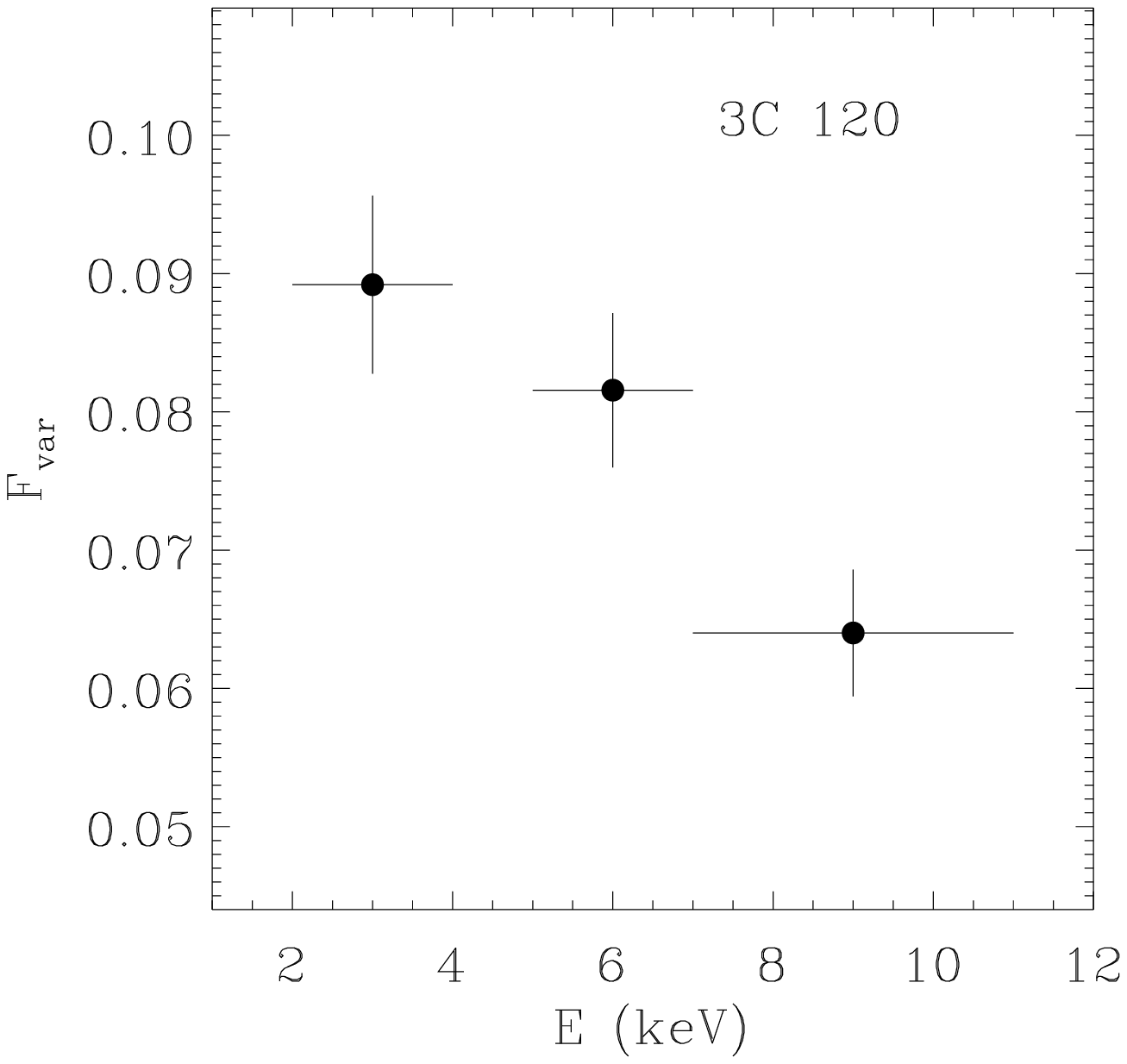,height=6.cm,width=10.5cm,%
bbllx=-140pt,bblly=69pt,bburx=420pt,bbury=430pt,angle=0,clip=}
\caption{X-ray light curves of 3C~120 
from \rxte\ PCA observations in (top to bottom) the 2-5 keV, 5-7 keV, 7-11 keV bands. 
Time bins are 5760 s ($\sim$ one orbit). The bottom plot shows the fractional
variability parameter versus the energy.  
\label{figure:fvar120B}}
\end{figure*}

More interesting results are obtained by computing the fractional
variability parameter on selected energy bands. For both 3C~390.3 and
3C~120 we used light curves in three energy bands, 2-5 keV (soft
band), 5-7 keV (medium band) and 7-11 keV (hard band), and calculated
$F_{\rm var}$ for all of them.  All bands show significant
variability. The mean count rate $\langle rate\rangle$, 
and the fractional variability parameter $F_{\rm
var}$ calculated using time bins of 1440~s ($\sim$ a quarter of an
orbit) for the light curves of
3C~390.3 and 3C~120 in the soft, medium and hard band, are given in
Table~3. For both objects, we found that the amplitude of variability
decreases with increasing energy band, with 3C~390.3 showing a
fractional variability of 31\% in the soft band, and 25\% in the
hard band, while 3C~120 has $F_{\rm var}$ decreasing from 9\% in
the soft band to 6.4\% in the hard band.  These results are displayed
in Figure~\ref{figure:fvar390B} and Figure~\ref{figure:fvar120B},
along with the light curves in each of the three energy bands.  

We have carried out a $\chi^2$ test to
verify the hypothesis that the $F_{\rm var}$ {\it vs} energy band trend
is consistent with a constant. Both sources show evidence of statistically
significant variability of $F_{\rm var}$ {\it vs} energy:
for 3C~390.3 we obtained $\chi^2=5.2$ for 2 degrees
of freedom, with a chance probability $P_{\chi^2}\sim 
7\%$; for 3C~120 $\chi^2=12.9$, $P_{\chi^2}\ll 1\%$.

To quantify the degree of linear correlation between $F_{\rm var}$ and
the energy band (and all the other possible correlations thereafter),
we calculate the linear correlation coefficient {\it r} and compute
the chance probability $P_c(r;N)$ that a random sample of $N$
uncorrelated pairs of measurements would yield a linear correlation
coefficient equal or larger than $|r|$; if this chance probability is
small, the two quantities are likely to be correlated.  For 3C~390.3
we found $r=-0.999$ and $P_c\sim 2\%$, and for 3C~120 $r=-0.975$ and
$P_c\sim 10\%$, which confirm the presence of an anti-correlation between
$F_{\rm var}$ and the energy band.  The results of the $\chi^2$ test
and correlation analysis are reported in Table 3.

\begin{figure*}
\psfig{figure=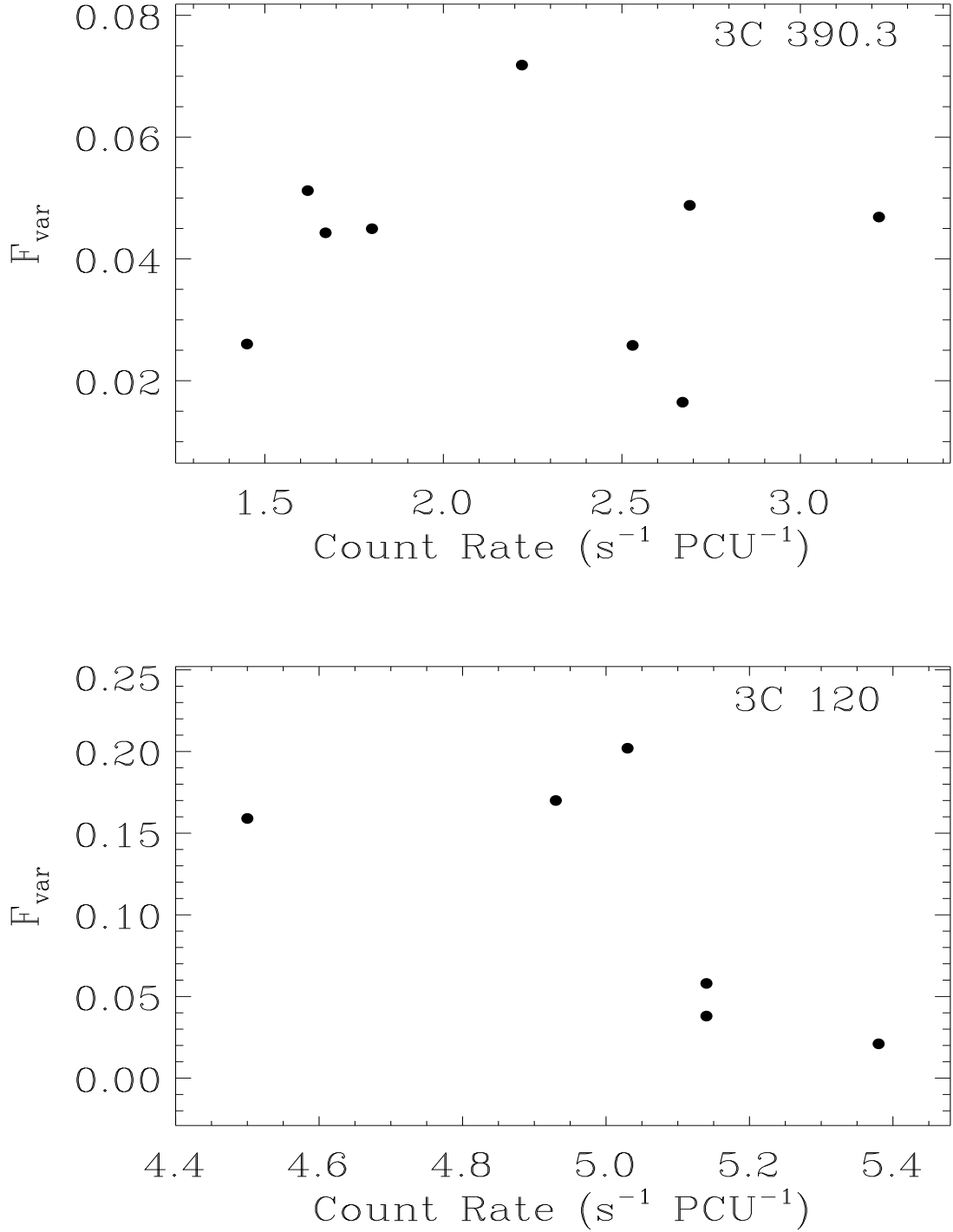,height=19cm,width=15cm,%
bbllx=41pt,bblly=69pt,bburx=450pt,bbury=654pt,angle=0,clip=}
\caption{$F_{\rm var}$ versus the 2--11 keV count rate, for 3C~390.3
(top panel) and 3C~120 (bottom panel). Time bins are approximately
equal to one \rxte\ orbit. The error bars are smaller than the
symbol size.  
\label{figure:var-cts}}
\end{figure*}

To examine the evolution of the timing properties of 3C~390.3 and
3C~120 further, we split the two long-term light curves, binned at
5760~s, into evenly sampled sub-intervals of nearly four days for
3C~390.3 and six days for 3C~120 (even sampling is important because
$F_{\rm var}$ depends strongly on the duration of the data-train). The
reason for choosing shorter intervals for 3C~390.3 is that we want to
measure the short-term variability (and its trend versus the local
count rate) without the bias of the long-term variability trend.  We
then calculated $F_{\rm var}$ over each interval and found that
3C~390.3 shows no evidence for significant variability
($\chi^2_{\rm red}=1.03$, $P_{\chi^2}\sim 40\%$)
nor correlation between the variability and
count rate ($r=-0.05,~ P_c\simeq 90\%$), while the fractional
variability of 3C~120 is statistically variable
($\chi^2_{\rm red}=34.9$, $P_{\chi^2}\ll 1\%$) and marginally
anticorrelated with the local
average count rate ($r=-0.67,~ P_c\simeq 15\%$). In any case, 
Figure~\ref{figure:var-cts} (bottom panel)
shows that the flux variability parameter of 3C~120 is significantly
smaller for count rate values above a certain threshold. 

\subsection{X-Ray Colors}

\begin{figure*}
\psfig{figure=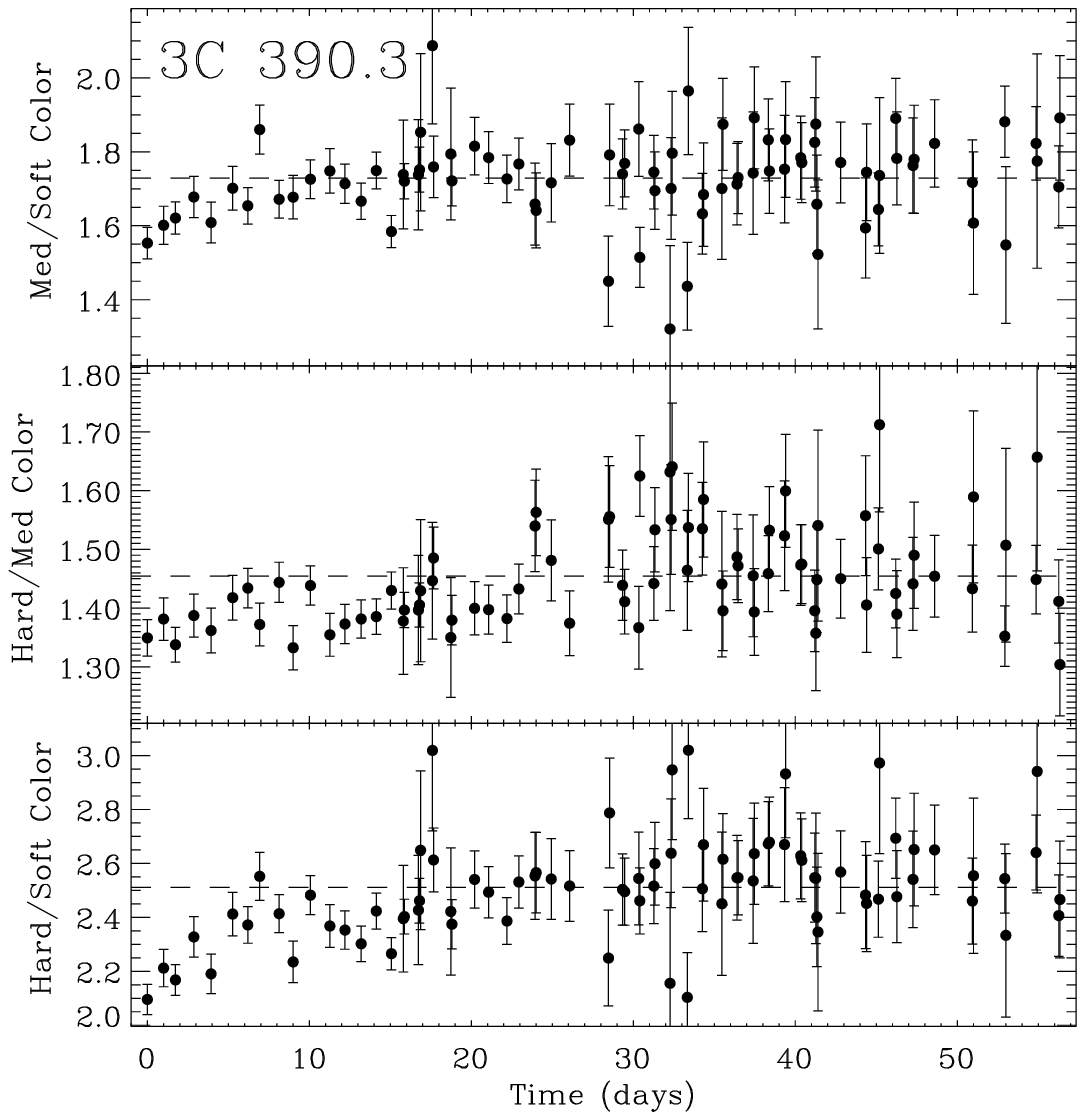,height=11.cm,width=13cm,%
bbllx=80pt,bblly=44pt,bburx=396pt,bbury=372pt,angle=0,clip=}
\psfig{figure=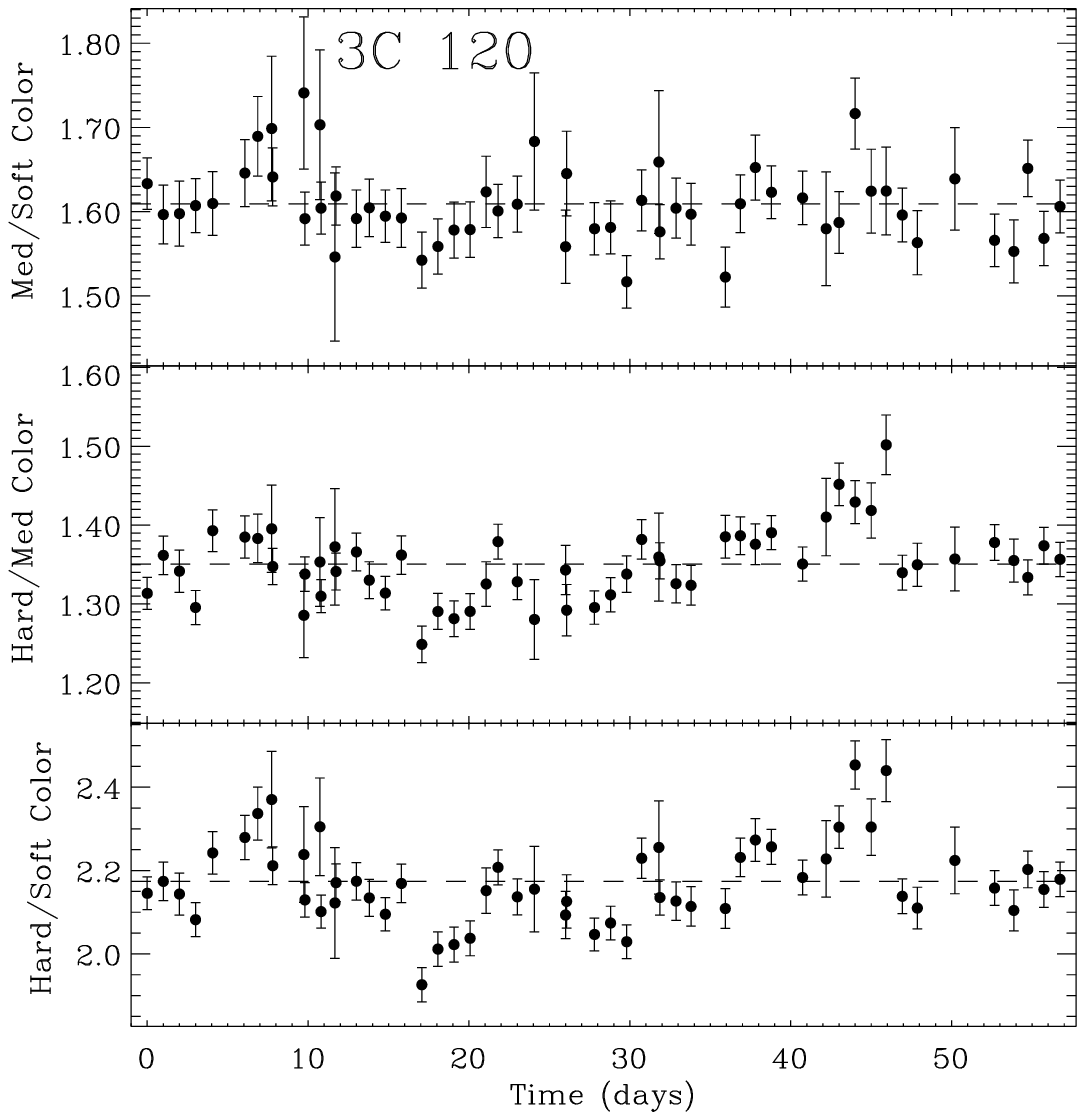,height=11.cm,width=13cm,%
bbllx=80pt,bblly=44pt,bburx=396pt,bbury=372pt,angle=0,clip=}
\caption{Hardness ratio light curve for 3C~390.3 (top panel) and 3C~120 (bottom panel).
Time bins are 5760 s ($\sim$ one \rxte\ orbit).   
\label{figure:HRt}}
\end{figure*}

X-ray colors provide a simple, model-independent way of investigating
spectral variability.  We defined three colors as ratios of count
rates in different bands: ${\rm Hard/Soft}=F_{7-11}/F_{2-5}$, ${\rm
Hard/Med}=F_{7-11}/F_{5-7}$, and ${\rm Med/Soft}=F_{5-7}/F_{2-5}$.
Figure~\ref{figure:HRt} shows ${\rm Med/Soft}$, ${\rm Hard/Med}$
and ${\rm Hard/Soft}$ plotted versus time for both 3C~390.3 and 3C~120.  
According to a $\chi^2$ test, all the colors with the only exception of 
the Med/Soft color in 3C~120 exhibit statistically significant
variations, with a higher variability for the Hard/Soft ratio. For the
Med/Soft color of 3C~390.3 we found $\chi^2=108$ corresponding to a 
chance probability $< 2\%$ for 81 d.o.f; $\chi^2=149$ and 241 (with
P$_{\chi^2}\ll1\%$ in both cases) for Hard/Med and Hard/Soft, respectively.
For the Med/Soft color of 3C~120 we found $\chi^2=58.3$ corresponding to a 
chance probability of $\sim 32\%$ for 54 d.o.f;
$\chi^2=147$ for Hard/Med and $\chi^2=210$  for 
Hard/Soft with P$_{\chi^2}\ll1\%$  in both cases. The
presence of spectral variability is also clearly demonstrated in the
plots of the color against the count rate in Figure~\ref{figure:HRt}.

\begin{figure*}
\psfig{figure=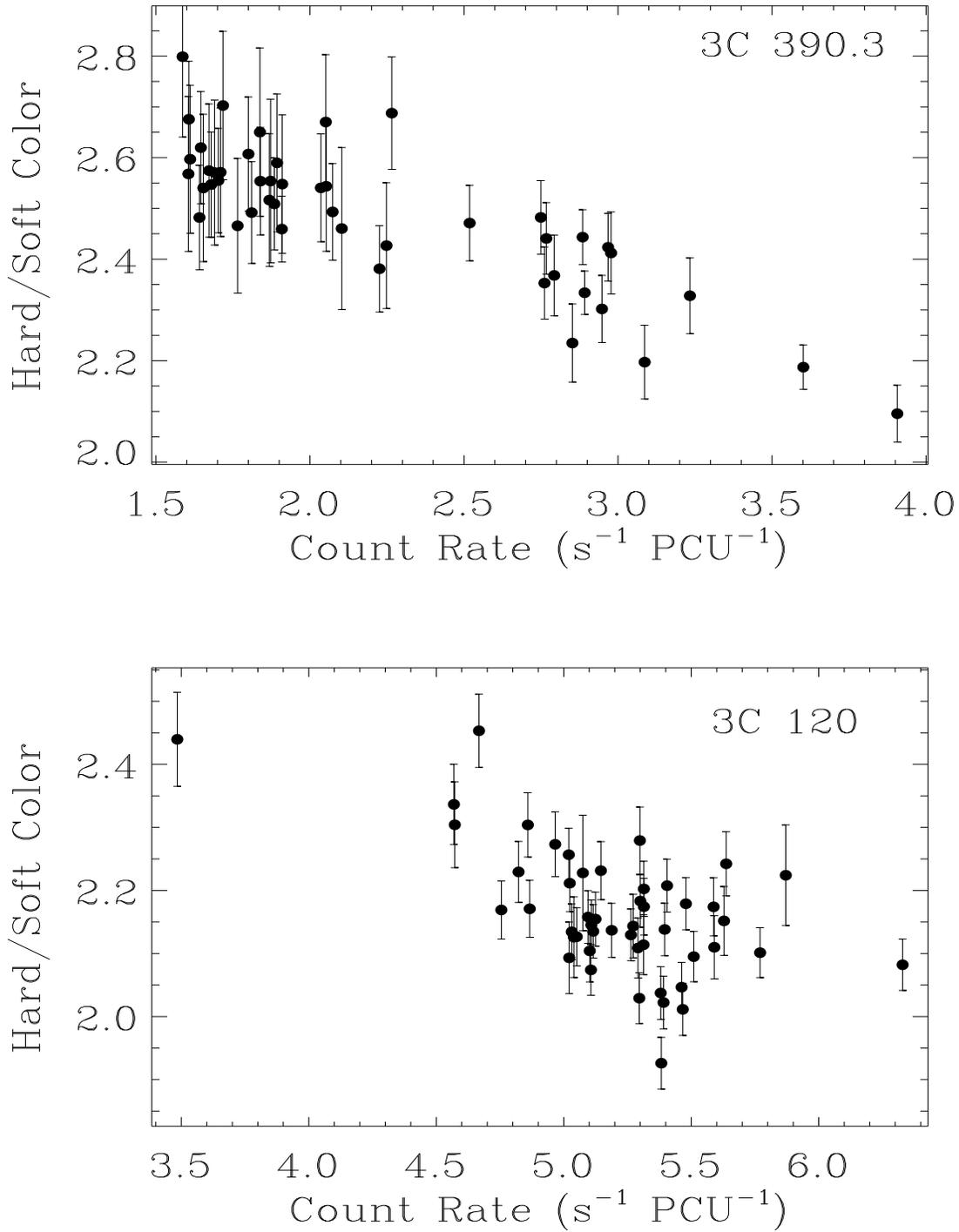,height=19cm,width=15cm,%
bbllx=50pt,bblly=68pt,bburx=460pt,bbury=656pt,angle=0,clip=}
\caption{Hard/Soft of 3C~390.3 (top) 3C~120 (bottom panel) plotted 
against the 2--11 keV count rate. Time bins are $\sim$ one orbit.
\label{figure:HR-cts}}
\end{figure*}

Figure~\ref{figure:HR-cts} shows the X-ray color ($F_{7-11}/F_{2-5}$)
of 3C~390.3 (top panel) and 3C~120 (bottom panel) plotted against the
average count rate in the 2-5 keV + 7-11 keV energy band. Both sources
show a strong anticorrelation between colors and count rates:
the linear correlation coefficients are $r=-0.84$ for 3C~390.3
and $r=-0.59$ for 3C~120; the probability that uncorrelated data
would give a linear correlation coefficient larger than the values found
in either case is much smaller than 1\%.
With a linear least square fit we found the following slopes:
$(-19.8\pm1.9)\times 10^{-2}$ for 3C~390.3 and $(-1.4\pm0.3)\times
10^{-2}$ for 3C~120.

\begin{figure*}
\psfig{figure=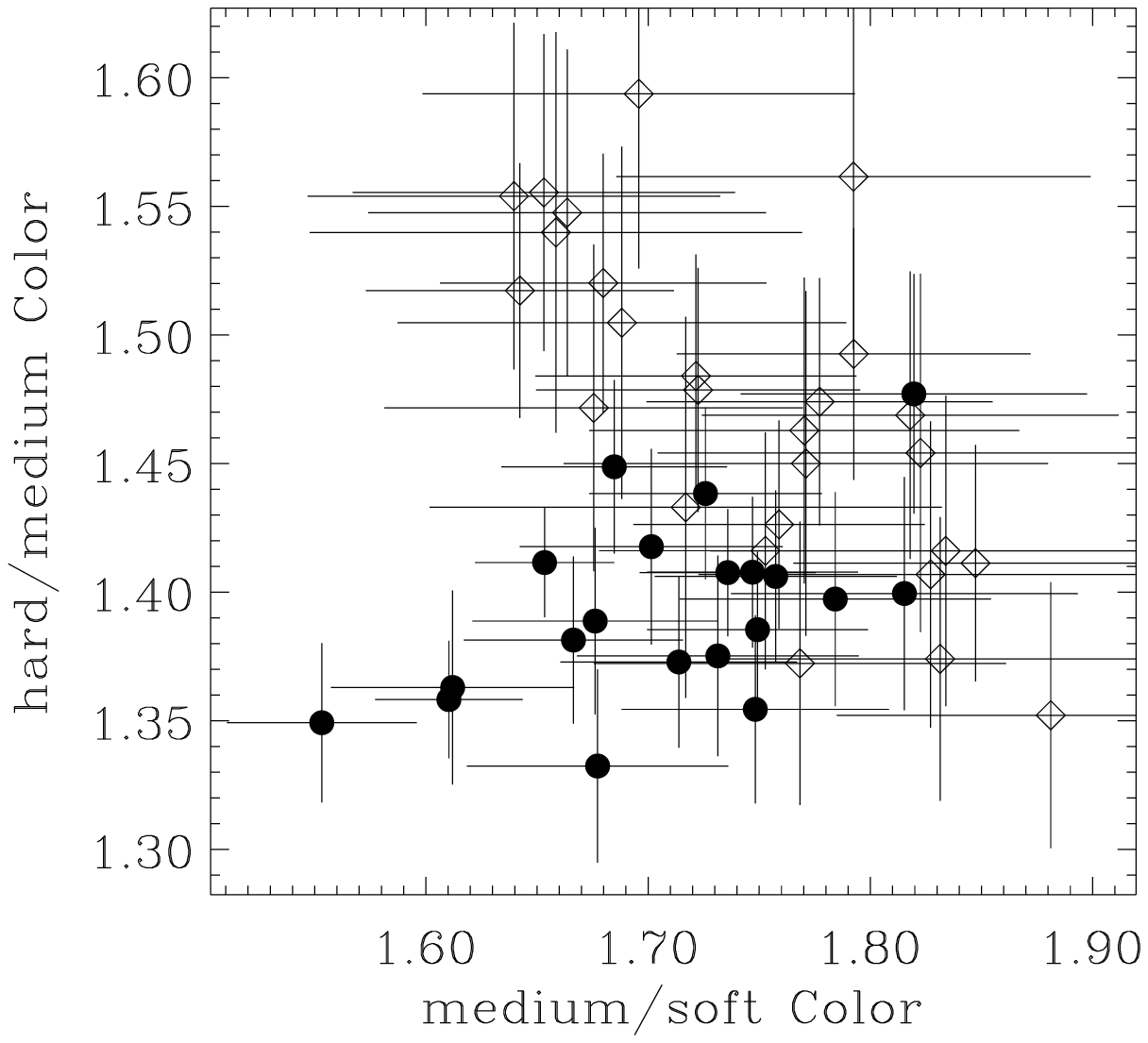,height=11.cm,width=13cm,%
bbllx=50pt,bblly=90pt,bburx=405pt,bbury=420pt,angle=0,clip=}
\psfig{figure=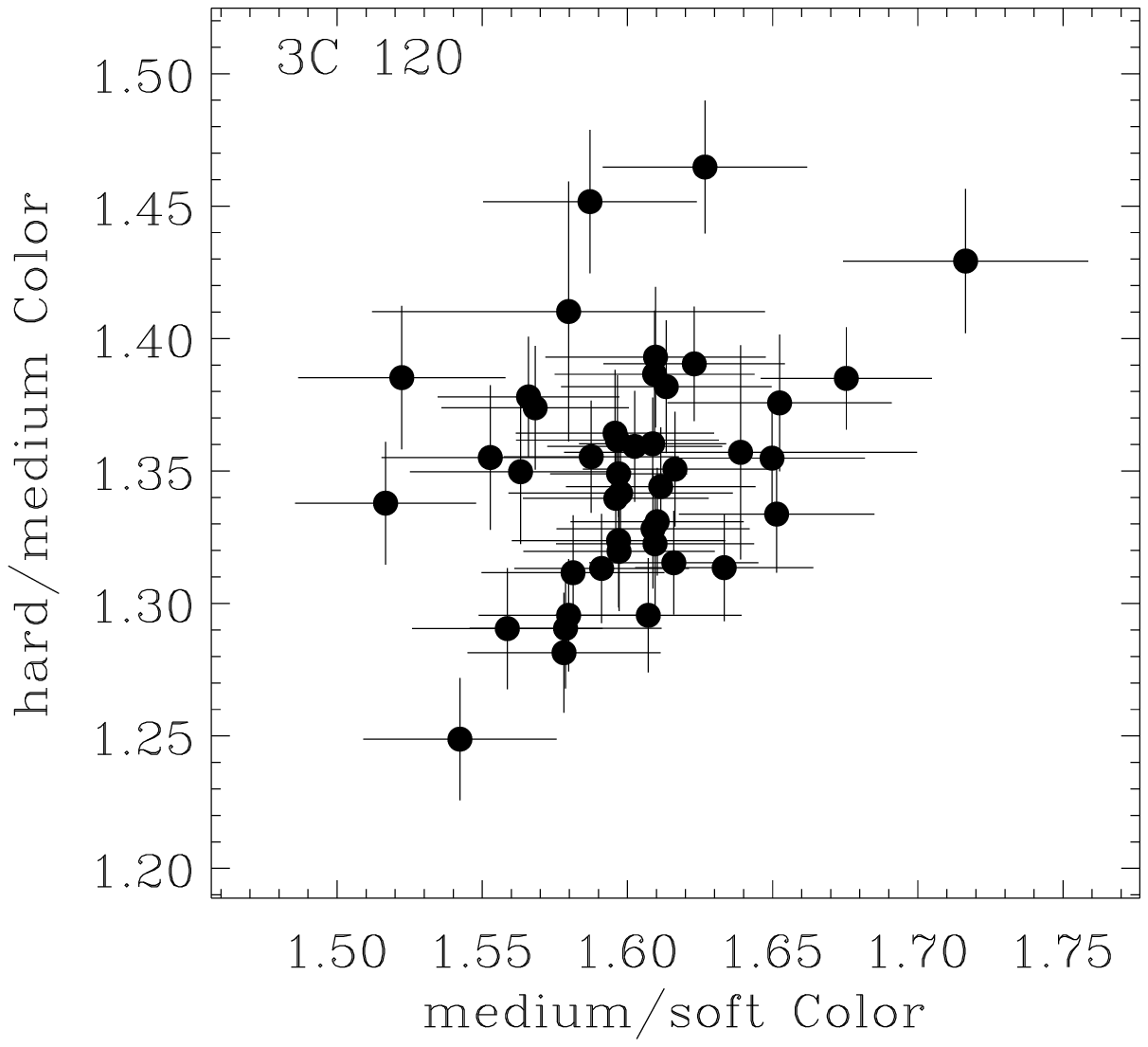,height=11.cm,width=13cm,%
bbllx=50pt,bblly=90pt,bburx=405pt,bbury=420pt,angle=0,clip=}
\caption{Color-color diagrams for 3C~390.3 (top panel) and 3C~120 (bottom panel).
Time bins are 1 day. Filled dots refers to the first half of the  3C~390.3 light
curve and the open diamonds to the second half.
\label{figure:color}}
\end{figure*}

Having selected three continuum energy bands, we can also construct
color-color diagrams. The main purposes for this analysis is to show:
1) that the two sources, similar in many aspects, behave
differently, as already pointed out by the structure function analysis, and,
more importantly, 2) that 3C~390.3 show a distinct bimodal behavior during the
monitoring campaign (see text below).
Figure~\ref{figure:color} shows the color-color
plots of 3C~390.3 (top panel) and 3C~120 (bottom panel). While in the
latter the points are crowded in the central region with evidence
for a weak positive correlation ($r=0.32,~ P_c\simeq 3\%$),
the color-color diagram of 3C~390.3 looks more complex. To investigate
the possible correspondence of the positions in the color-color plot
with the evolution of the light curve, we have plotted with filled
dots the values corresponding to the first half of the light curve
(i.e. during the smooth decay phase) and with open diamonds the values
corresponding to the second half of the light curve (i.e. during the
phase of nearly stationary flux).  The color-color plot clearly
indicates the presence of a bimodal trend: during the decay phase of the 
light curve the points
are positively correlated ($r=0.51,~ P_c=2\%$), 
while during the stationary phase there is
a strong inverse correlation ($r=-0.72,~ P_c\ll 1\%$)
between Hard/Med and Med/Soft. In fact the
behavior of 3C~390.3 in color-color space is reminiscent of the
Z-shaped tracks of some low-mass X-ray binaries (the ``Z sources'';
Hasinger \& van der Klis 1989; see also van der Klis 1995 for a recent
review). In the case of 3C~390.3, not all three branches of the
Z-shaped track are traced, only the horizontal and normal branches.

\section{Spectral Analysis}

The presence of pronounced spectral variability, as shown by the color
analysis in the previous section for both 3C~390.3 and 3C~120,
indicates that a time-averaged spectral analysis might not be the most
appropriate tool for the diagnostic of physical conditions in the
inner regions of 3C~390.3 and 3C~120. However, the limited
signal-to-noise ratio of HEXTE data allows only a time-averaged
spectral analysis.  Moreover, even if the spectral index of the
``primary" power-law spectrum does vary with luminosity, the spectral
features resulting from the reprocessing of the primary X-rays in the
accretion disk should not be significantly affected by these
spectral-index fluctuations, especially since the the range of
variation turns out to be small (see \S4.2, below). The primary cause
of fluctuations in the reprocessing features is the variation of the
primary power-law intensity, which itself spans a range of
approximately a factor of 2. Under these conditions, we expect that
the spectrum of reprocessed X-rays varies mainly in intensity and
not in any other way (see, for example, Ballantyne \& Ross 2002). Thus
we use the time-averaged spectrum primarily to evaluate reprocessing
models and not models for the primary continuum.  Therefore, we first
performed a time-averaged analysis using the PCA and HEXTE data
together, in the attempt to constrain the Compton ``reflection''
component. In a second pass we analyzed time-resolved PCA spectra of
3C~390.3 to study the evolution of the spectral index and possible
variations of the \feka\ line flux over the course of the large,
systematic decline displayed during the first half of the monitoring
period.

The spectral analysis of PCA and HEXTE data was performed using the
{\tt XSPEC v.11} software package (Arnaud 1996). We used PCA response
matrices and effective area curves created specifically for the
individual observations by the program {\tt pcarsp}, taking into
account the evolution of the detector properties. To fit the HEXTE
spectra we used response matrices and effective area curves created on
1997 March 20. The PCA spectra (with the appropriate response matrices)
were first extracted selecting several short time intervals to take into
account the different number of PCUs at work and any possible change
in calibration; subsequently, all the spectra were combined with 
{\tt addspec}.
All the spectra were rebinned so that each bin
contained enough counts for the $\chi^2$ statistic to be valid. Fits
were performed in the energy ranges 4--20 keV (PCA), 20-50 keV (HEXTE),
where the signal-to-noise ratio is the highest. Joint fits to
PCA+HEXTE spectra were carried out allowing the normalization factors
for each instrument to be a free parameter.

\subsection{Time-Averaged Spectra}

\begin{figure*}
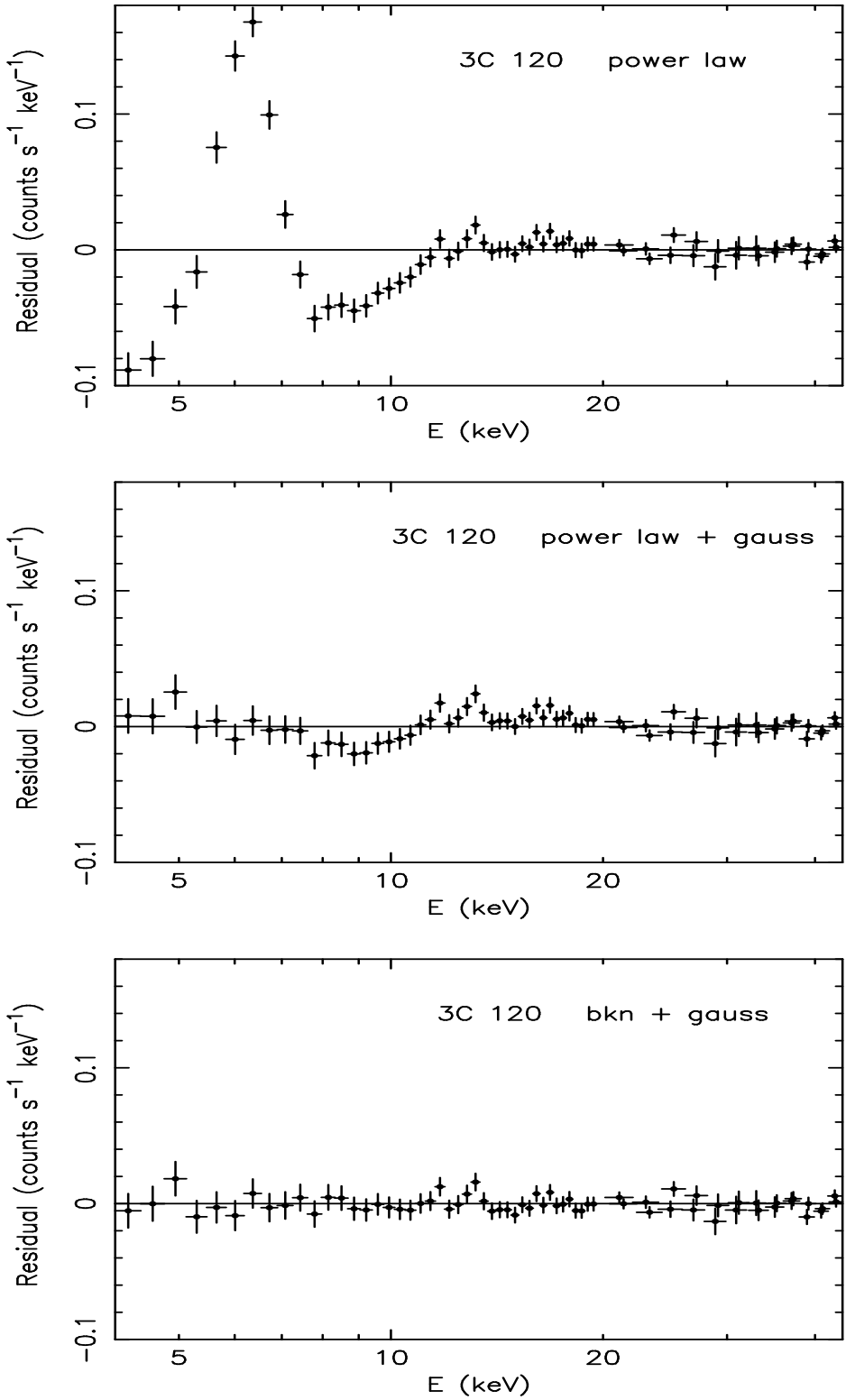

\psfig{figure=fig9a.ps,height=7.cm,width=13cm,%
bbllx=539pt,bblly=643pt,bburx=117pt,bbury=91pt,angle=-90,clip=}
\psfig{figure=fig9b.ps,height=7.cm,width=13cm,%
bbllx=539pt,bblly=643pt,bburx=117pt,bbury=91pt,angle=-90,clip=}
\psfig{figure=fig9c.ps,height=7.cm,width=13cm,%
bbllx=539pt,bblly=643pt,bburx=117pt,bbury=91pt,angle=-90,clip=}
\caption{Spectral fit residuals to the PCA+HEXTE spectrum of 3C~120 for simple power 
law (top panel), a power law plus a Gaussian component at
6.4 keV (middle panel), a broken power law plus a Gaussian line at 6.4 keV (bottom
panel).   
\label{figure:120res}}
\end{figure*}

We first fitted the combined PCA+HEXTE spectra of 3C~390.3 and 3C~120
with a simple power-law with Galactic absorption. The very poor fits
($\chi^2=420.6/49~dof$ and $\chi^2=885.8/67~dof$, for 3C~390.3 and
3C~120 respectively), clearly indicate the need for a different
continuum model or, at least, of an additional component. In fact,
adding a Gaussian line improved the spectral fits significantly. This
is clearly evident in Figure~\ref{figure:120res}, where we plot the
residuals to the PCA+HEXTE spectrum of 3C~120 for a simple power-law
model (top panel), a power-law plus a Gaussian component at 6.4 keV
(middle panel), a broken power-law plus a Gaussian line at 6.4 keV
(bottom panel). Similar results are obtained for 3C~390.3.

\begin{figure*}
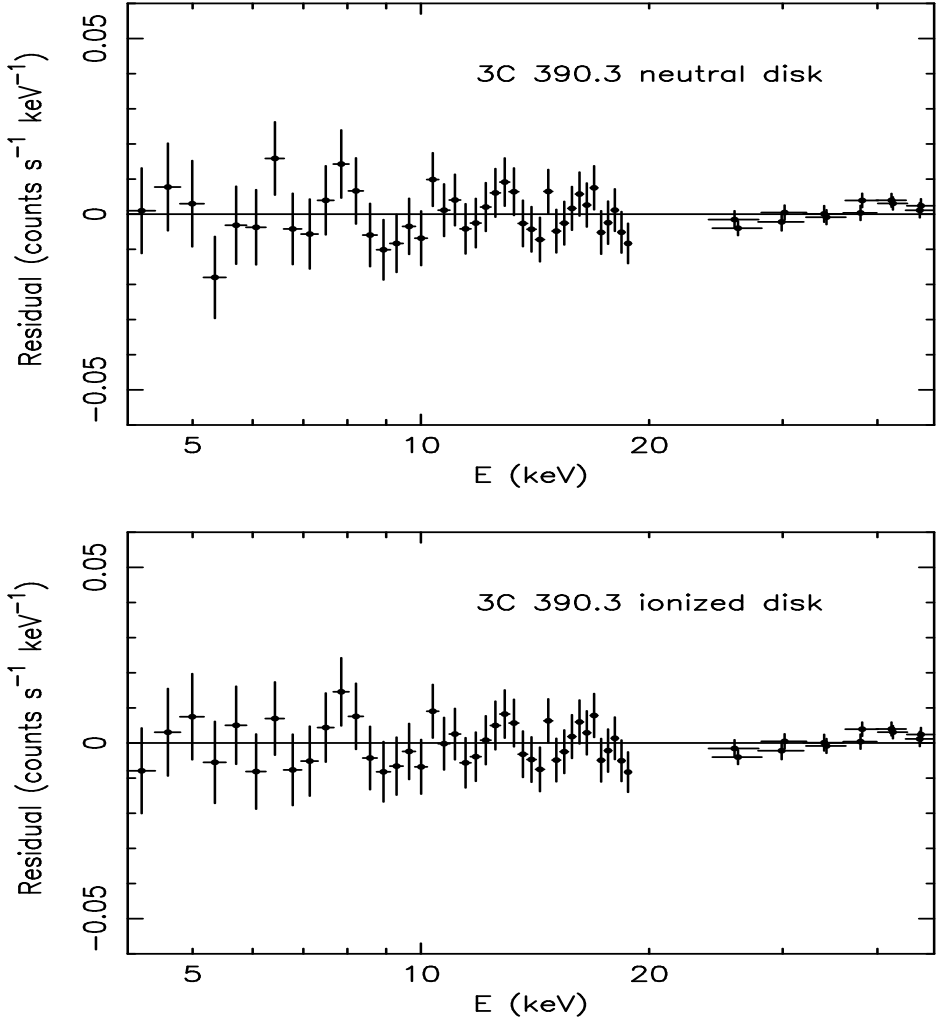

\psfig{figure=fig10a.ps,height=7.cm,width=13cm,%
bbllx=539pt,bblly=643pt,bburx=117pt,bbury=91pt,angle=-90,clip=}
\psfig{figure=fig10b.ps,height=7.cm,width=13cm,%
bbllx=539pt,bblly=643pt,bburx=117pt,bbury=91pt,angle=-90,clip=}

\caption{Spectral fit residuals to the PCA+HEXTE spectrum of 3C~390.3 for the
{\tt pexrav} model (top panel), and for the constant-density ionized-disk model
(bottom panel).   
\label{figure:390res}}
\end{figure*}

The rather weak reflection component usually found in BLRGs can be
interpreted either in terms of a truncation of the innermost parts of
the optically-thick accretion disk (thought to be the main
``reflector") or, as recently suggested by Ballantyne et al. (2002),
as a result of reprocessing in an {\it ionized} accretion disk.  For
this reason, as models of the continuum, apart from a single power-law
and a broken power-law, we tried a power-law plus its Compton
reflection, implemented in {\tt XSPEC} as {\tt pexrav} (Magdziarz \&
Zdziarski 1995) and the constant-density ionized-disk reflection model
(Ross \& Fabian 1993, Ballantyne et al. 2001) available as a
table model in {\tt XSPEC v11.2}. 
Figure~\ref{figure:390res} shows the residuals of the PCA+HEXTE spectrum of 
3C~390.3 for the {\tt pexrav} model (top panel) and for the constant-density 
ionized-disk reflection model (bottom panel), with an additional Gaussian
line at 6.4 keV. In the latter model, the
accretion disk is approximated as a slab of gas with constant density
($n_{\rm H}=10^{15}{\rm~cm^{-3}}$) and solar abundances, which is
illuminated by X-rays with a flux $F_{\rm x}$ (extending from 1 eV to 100 keV)
and a power-law spectrum
of photon index $\Gamma$. The reflected spectrum is scaled according to 
the solid angle of the absorber ($R\equiv\Omega/2\pi$),
and then added to the incident spectrum to
give the final observed spectrum. The structure of the reflected
spectrum is determined by the ionization parameter $\xi=4\pi F_{\rm
x}/n_{\rm H}$. This model predicts the \feka\ emission line and the
spectral features at low energy (emission lines and recombination
continuum).  An additional Gaussian line at 6.4 keV (in the source rest frame) is
needed by all the continuum models to account for the strong
excess around 6.4 keV.  The results of fitting the above models are
summarized in Tables~4 and 5.

For both sources the best-fit is obtained using broken power-law
models: $\Delta\chi^2$ ranges between 11 (for 3C~390.3) and 60 (for
3C~120), which is significant at more than 99\% confidence for two
additional free parameters, with respect to the fit with a single
power-law.  However, since a broken power-law model is an {\it ad hoc}
phenomenological model, it is difficult to draw information on the
physical conditions of the accretion flow from spectral fits with this
model. More useful information can be derived by comparing the
spectral fits of models assuming reflection from a neutral and from an
ionized accretion disk, respectively. The first important result is
that the ionized disk reflection model gives a better description of
the time-averaged spectra of 3C~390.3 and 3C~120 (fits with {\tt
pexrav}, which has an additional free parameter with respect to the
ionized disk model, result in a significant increase of the total
$\chi^2$). However, for both sources the inferred ionization parameter
is not high, $\xi\simeq 300$, which is nearly a factor 10 lower than
the value derived for 3C~120 by Ballantyne et al. (2002) using a 50 ks
observation from ASCA, with the reflection fraction fixed at 1. We
tried to fit the ionized disk model with the reflection fraction fixed
at 1 to the \rxte\ data of 3C~390.3 and 3C~120, but the resulting
$\chi^2$ are not acceptable.  The reflected fraction (i.e.,
$\Omega/2\pi$) is negligible according to both the neutral and the
ionized disk models (consistent with zero for 3C~390.3).  To assess
whether the results from the time-averaged spectral analysis are
reliable (in view of the spectral variability shown by the X-ray color
analysis; see $\S3.3$), we have fitted nine time-resolved PCA spectra
plus the average HEXTE spectrum of 3C~390.3 with the neutral and
ionized-disk models. The main results are the following: 1) the
spectra are better fitted by an ionized-disk in six cases out of nine;
2) the spectral parameters, with the exception of the folding energy
for {\tt pexrav} do not vary significantly and are basically
consistent with the time-averaged values: the reflection is low for
both models and the parameter $\xi$ in the ionized-disk model is
always small, but not consistent with zero.

\begin{figure*}
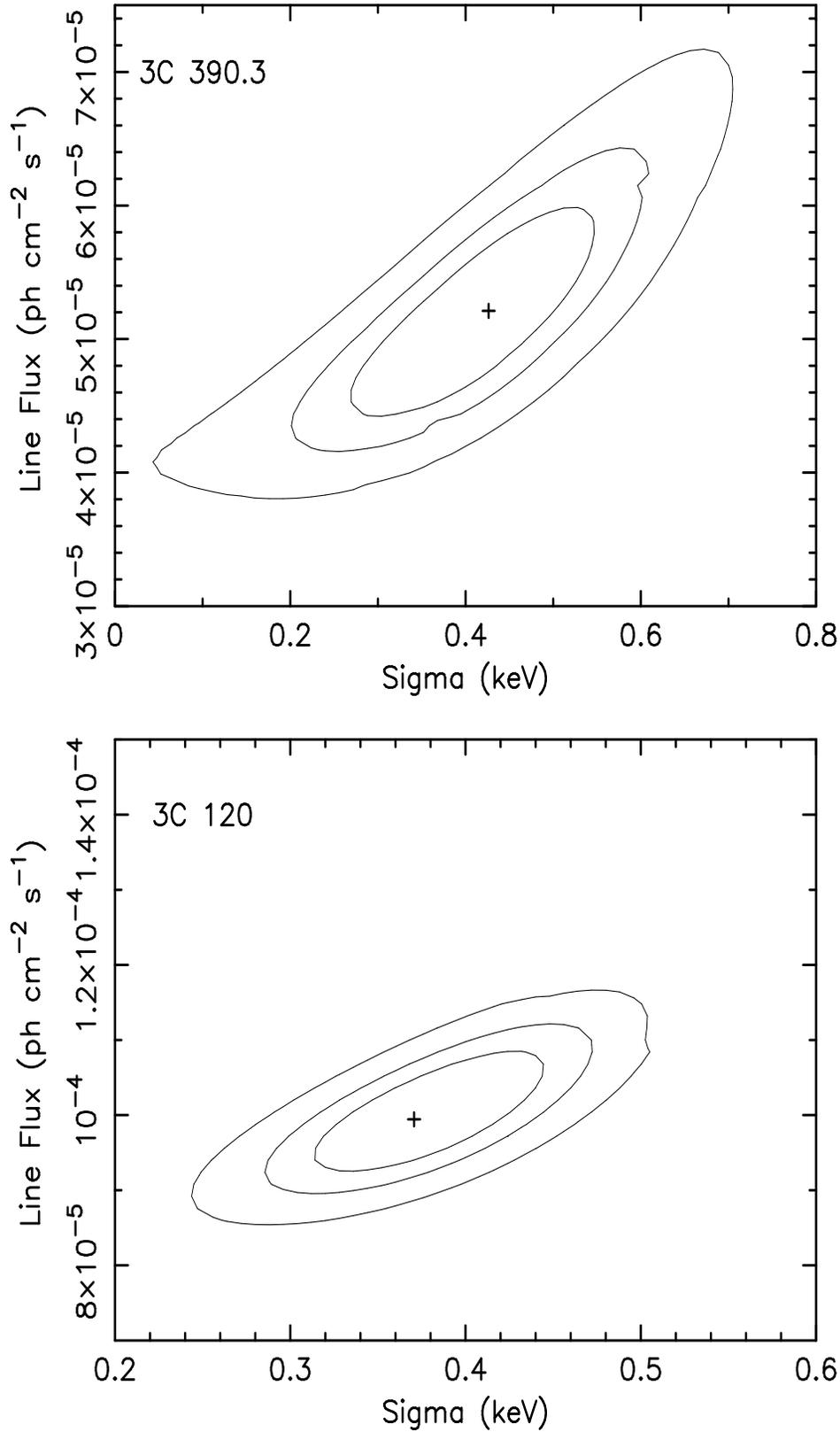

\psfig{figure=fig11a.ps,height=11.cm,width=13cm,%
bbllx=535pt,bblly=648pt,bburx=123pt,bbury=86pt,angle=-90,clip=}
\psfig{figure=fig11b.ps,height=11.cm,width=13cm,%
bbllx=535pt,bblly=648pt,bburx=123pt,bbury=86pt,angle=-90,clip=}
\caption{Confidence contours ($68\%$,$90\%$, and $99\%$) in the $\sigma$ - 
$I_{\rm Fe K\alpha}$ plane. The EW of the line is directly proportional
to its photon flux. The conversion factor can be derived from
the best-fitting model parameters  in Table~5.
\label{figure:contour}}
\end{figure*}

While the reflection component is weak, a strong iron line  is required
in all the spectral fits (if the spectral model used to fit the continuum is 
not the ionized-disk). The \feka\ line was modeled as a Gaussian of
energy dispersion $\sigma$, and intensity $I_{\rm Fe K\alpha}$. Due to
the low spectral resolution of the PCA, we did not use more
sophisticated models for the line profile.  Since HEXTE spectra do not
extend below 20 keV, for both objects we used only the PCA data, for
which a power-law + Gaussian line give acceptable fits of the
spectra. Both 3C~390.3 and 3C~120 have relatively broad ($\sigma\simeq
0.4$ keV), relatively strong ($I_{\rm
Fe~K\alpha}=5-10\times10^{-5}{\rm~photons~s^{-1}~cm^{-2}}$) iron
lines, with centroids consistent with the value of ``cold" \feka\ .
On the other hand, using the ionized-disk model to fit the continuum, 
the parameters of the additional
Gaussian line have much lower values: $\sigma\simeq
0.2$ keV, $I_{\rm Fe~K\alpha}=3\times10^{-5}{\rm~photons~s^{-1}~cm^{-2}}$.
The reason for this discrepancy is that the ionized-disk model already
includes the \feka\ emission line; as a consequence, part of the excess around
6.4 keV is already accounted for by the reprocessing model. A natural question
that can be raised at this point is: why an additional Gaussian component is 
necessary also for the ionized-disk model? 
There are two possible explanations: 1) if the
line originates from an inner region close to the black hole, relativistic
effects become important, but they are not taken into account by the 
constant-density ionized-disk model available in {\tt XSPEC}; 
2) the iron line emission is complex, possibly made of two main components: 
a broad component originating from the inner part of the accretion disk, and a second
narrow one probably associated with a region distant from the central source. 
Such a narrow line component has been detected in several Seyfert 1 
galaxies with \chandra\ (e.g., Yaqoob et al. 2001, Kaspi et al 2001) and 
\xmm\ (e.g., Pounds et al. 2001, Reeves et al. 2001).
It also is worth noting that De Rosa et al. (2002) found that
the BeppoSAX spectrum of the Seyfert 1 galaxy NGC 7469 was best fitted
with the ionized-disk model and  a second narrow line component, in addition 
to that produced in the disk, was also required to fit 
the iron line profile.
Spectral parameters of the \feka\ line in 3C~390.3 and 3C~120 are
reported in Table~5 and contour plots of the energy dispersion
$\sigma$ and the \feka\ photon flux are shown in
Figure~\ref{figure:contour}, indicating that in both sources the
\feka\ is resolved at more than 99\% confidence level.  

\subsection{Time-Resolved Spectra}

Since the data consist of short snapshots spanning a long temporal
baseline, they are well suited for monitoring the spectral variability of
the sources and represent the ideal tool for the diagnostic of the
changing physical conditions of the matter around the supermassive
black hole.  A test of scenarios for the geometry of the accretion
flow is afforded by the correlated or delayed variability of the
\feka\ line flux with the X-ray continuum flux (in effect, X-ray
reverberation mapping), as different time delays are expected
depending on the location of the cold reprocessor with respect to the
X-ray source. If the \feka\ line originates from the outer regions of
an ADAF, the time delay between the line and the continuum should be
of the order of a few days or larger. Shorter delays ($\leq$ 1 day)
are expected if the line originates from the inner parts of a disk
extending further in, as in Seyferts. Finally, if the line comes from
the molecular torus, the delays will be of the order of several months
or longer.

\begin{figure*}
\psfig{figure=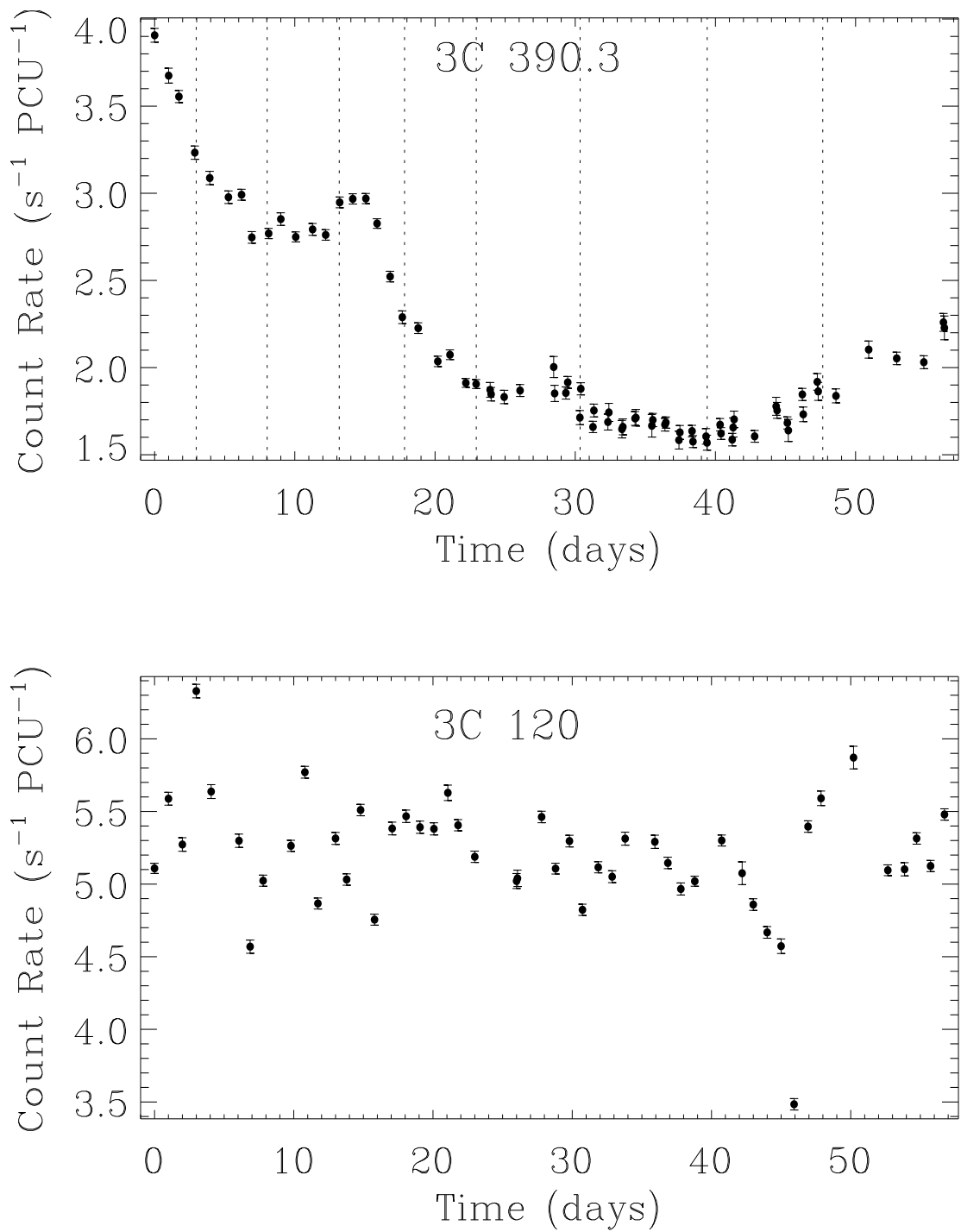,height=19cm,width=15cm,%
bbllx=44pt,bblly=64pt,bburx=449pt,bbury=543pt,angle=0,clip=}
\caption{X-ray light curves of 3C~390.3 (top panel) and 3C~120 (bottom
panel) from \rxte\ PCA observations in the 2--11 keV band.  Time bins
are 5760 s ($\sim$ one \rxte\ orbit). For the sake of clarity, only data
bins which were at least 10\% full are plotted here, although all the
data were used in the analysis. The dotted lines in the top panel
indicate the intervals used for the time-resolved spectral analysis.
\label{figure:lc1B}}
\end{figure*}

In an attempt to constrain the location of the gas responsible for the
\feka\ emission line in 3C~390.3, we performed an X-ray
``reverberation mapping'' study. This consists in the study of
possible delayed variations of the \feka\ line flux with the 2--10 keV
continuum.  Of our two sources, 3C~390.3 is the most suitable for
performing a study of the \feka\ line flux variability because of the
systematic trend of variability observed
(Figure~\ref{figure:lc1B}). We divided the light curve of
Figure~\ref{figure:lc1B} into nine intervals, with exposures ranging
between 10 ks and 25 ks, which represent a trade-off between the
necessity to isolate intervals with different but well defined average
count rates and the need to reach a signal-to-noise ratio high enough
for a meaningful spectral analysis.  The dates, exposure times and
mean fluxes in the nine selected intervals are listed in Table~6.

We fitted the 4--20 keV spectra with a model consisting of a power-law, 
absorbed by the Galactic absorbing column, plus a Gaussian \feka\
line with a rest energy fixed at 6.4 keV. 
We did not use the constant-density ionized model to fit the
continuum because, as explained above, although it already includes the \feka\
line, it does not allow us to measure line parameters.
The results of the fits are
summarized in Table~7.  Due to the combination of a short exposure and
a low count rate, the spectral data of the ninth interval have low
signal-to-noise ratio and therefore the spectral parameters inferred
from the fit are characterized by large errors. For this reason we
will not use these values in the following analysis.

\begin{figure*}
\psfig{figure=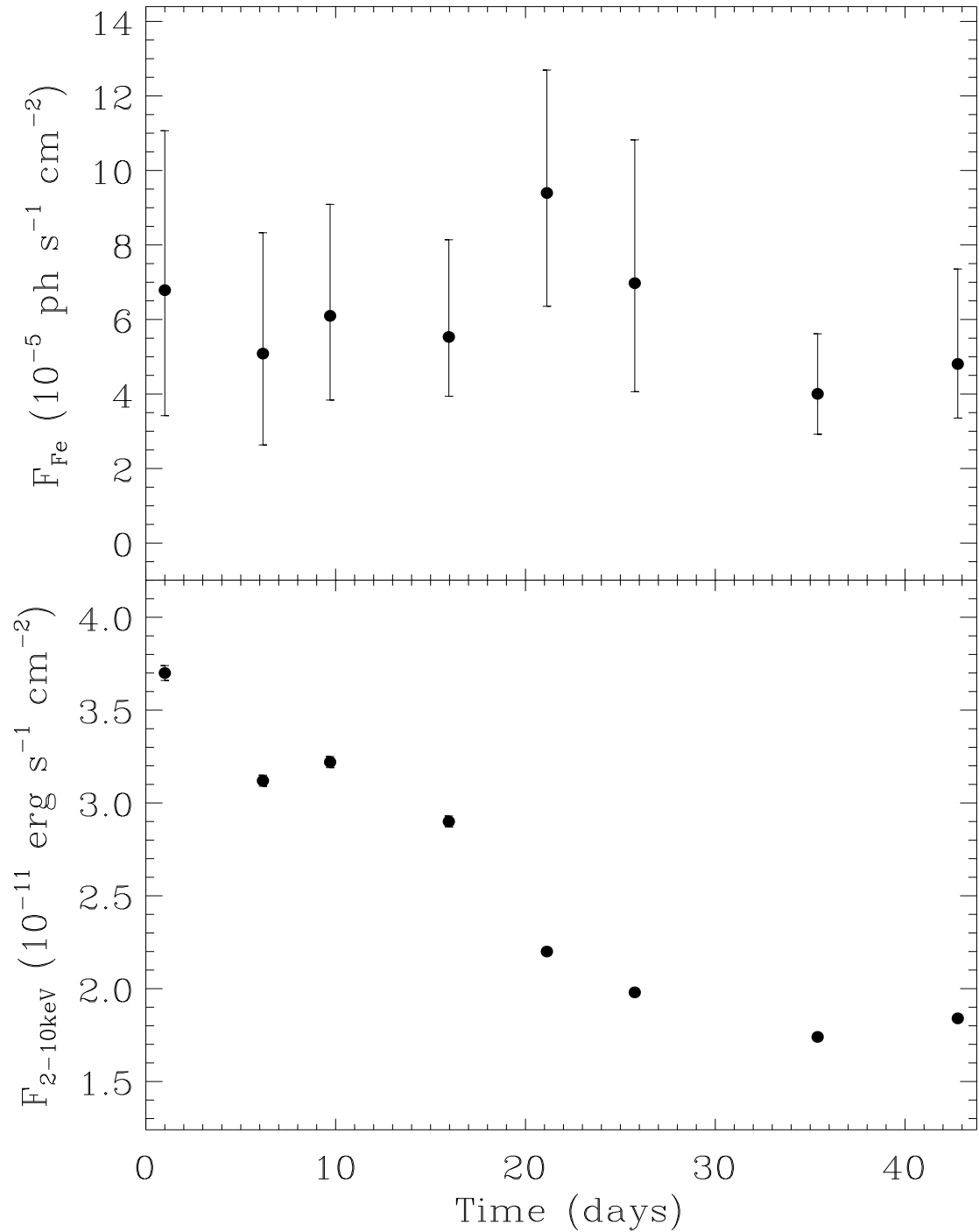,height=19cm,width=15cm,%
bbllx=82pt,bblly=62pt,bburx=488pt,bbury=598pt,angle=0,clip=}
\caption{Light curves of the 
\feka\ line flux ($10^{-5}{\rm~photons~s^{-1}~cm^{-1}}$; top panel)
and of the 2--10 keV flux ($10^{-11}{\rm~erg~s^{-1}~cm^{-1}}$; bottom panel).
The error bars correspond to 90\% confidence level.
\label{figure:reverb}}
\end{figure*}

The top panel of Figure~\ref{figure:reverb} shows the light curves of
the \feka\ line flux while the bottom panels shows the 2--10 keV flux
for comparison. Because of the large error bars, we can only speculate
about the implications of this diagram.  Neglecting the errors, one
might argue that the maximum of the \feka\ line flux light curve
appears to be shifted by $\sim 20$ days compared to the 2--10 keV flux
maximum, however the significance of the \feka\ line flux variability
is statistically questionable. A $\chi^2$ test shows that the \feka\
light curve is consistent with being constant ($\chi^2_r=0.6$ for
8~d.o.f.; $P_{\chi^2}\simeq80\%$), which discouraged us from attempting 
more sophisticated tests, such as cross-correlation analysis.

\begin{figure*}
\psfig{figure=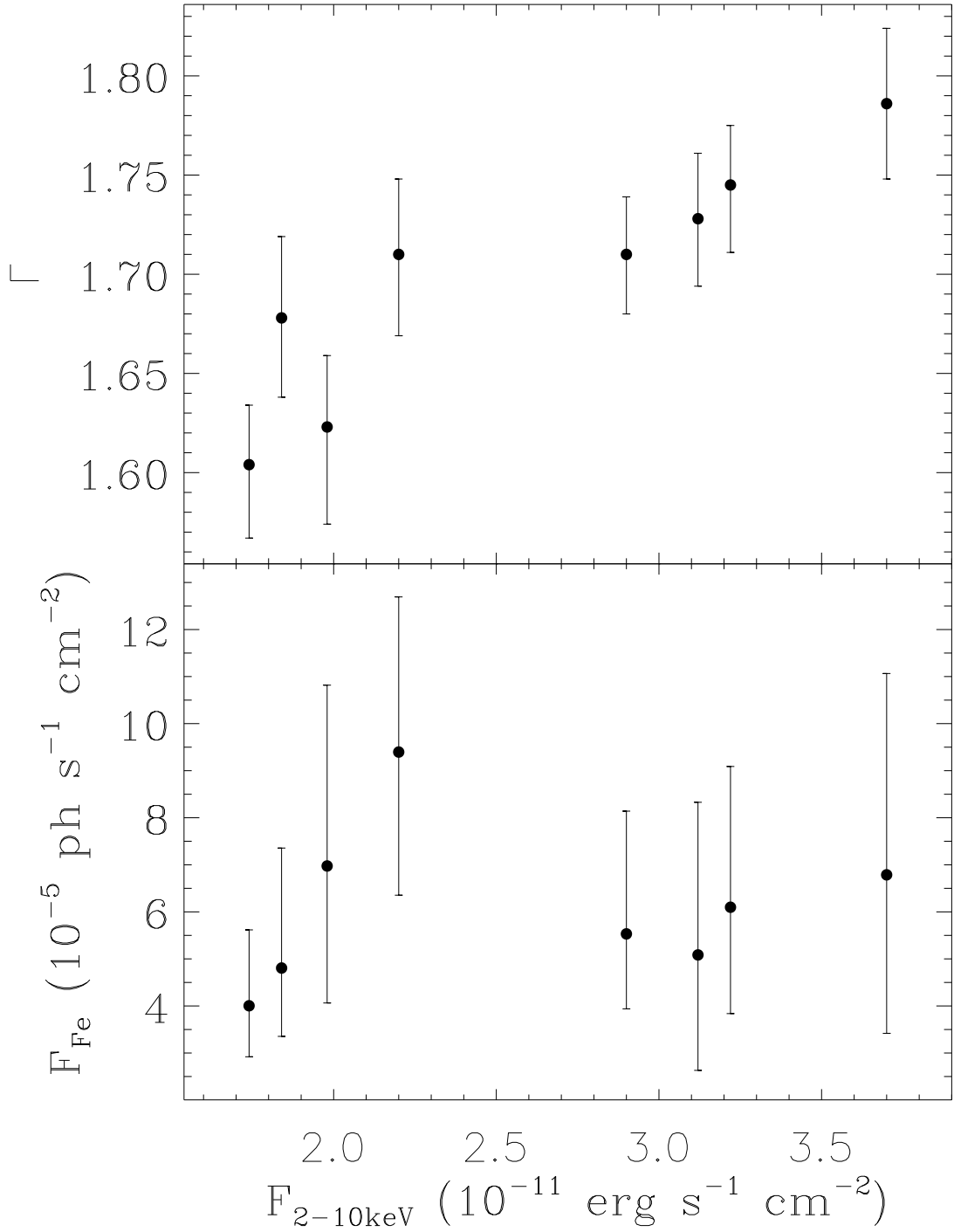,height=19cm,width=15cm,%
bbllx=50pt,bblly=50pt,bburx=488pt,bbury=598pt,angle=0,clip=}
\caption{Photon index $\Gamma$ in the range 4-20 keV (top panel) and 
\feka\ line flux ($10^{-5}{\rm~photons~s^{-1}~cm^{-1}}$; bottom panel)
plotted against the 2--10 keV flux ($10^{-11}{\rm~erg~s^{-1}~cm^{-1}}$).
\label{figure:specres2}}
\end{figure*}

Even the PCA sensitivity is not sufficient to obtain reliable results
from the time-resolved X-ray spectroscopy.  Nevertheless, valuable
information can be obtained by plotting spectral parameters against
the 2--10 keV flux. Figure~\ref{figure:specres2} shows the plots of the
photon index $\Gamma$ (top panel) and \feka\ line flux (bottom panel)
versus the 2--10 keV flux ($10^{-11}{\rm~erg~s^{-1}~cm^{-1}}$). The
main results are the following: the photon index increases with the
2--10 keV flux, with some indications of a flattening at higher fluxes.
The \feka\ line flux also increases as the 2--10 keV flux increases,
but only for a limited range of fluxes; at larger continuum flux
values the \feka\ line flux is roughly constant. 

Probably only \xmm\, with its superior sensitivity and the
possibility of long continuous observations will be able to achieve
significant results in this kind of analysis.

\section{Summary and Discussion}

We have presented a temporal and spectral analysis of two long ($\sim
60$ days) \rxte\ observations of the BLRGs 3C~390.3 and 3C~120
performed in May-July 1996 and January-March 1997, respectively.  The
first important result of the timing analysis is that the two sources
behave differently. This does not necessarily imply that
the two BLRGs are intrinsically different, only that they have been
caught during different facets of their temporal behavior.
The qualitative difference inferred by the visual
inspection of the light curves is quantified by the structure function
analysis, which shows that 3C~390.3 has a characteristic timescale of
$\sim$ 20 days (i.e., its decaying time), while no typical timescale
is found for 3C~120. Also the flux variability, quantified in terms of
the fractional variability parameter $F_{\rm var}$, is different:
3C~390.3 is significantly more variable than 3C~120, despite its jet
inclination angle inferred from radio observations ($19^\circ\leq
i\leq 33^\circ$) is larger than that of 3C~120 ($1^\circ\leq i\leq
14^\circ$). A straightforward interpretation of this first result is
that the X-ray variability we observe is not related to (or, at least,
not dominated by) the relativistic jet associated with the central
engine.  More specifically, in the case of 3C~390.3, significant
variations occur on a time scale of order a month, which is longer
than the typical variability time scale of blazars (e.g., Kataoka,
Takahashi, \& Wagner 2001).  In the case of 3C~120, there are clear
variations from day to day, but their amplitude is of order 10\%, for
which we offer the following plausible interpretation. We hypothesize
that the observed X-ray flux is a combination of variable emission
from the relativistic jet and steady emission from a source associated
with the inner accretion disk. If we take the jet emission to be
variable by a factor of 2 about its mean level, then the observed
variations at a level of 10\% imply that the contribution of the jet
emission to the total observed flux is only 5\%. The limited
contribution of the jet X-ray emission is also confirmed by the X-ray
spectrum, which shows a prominent \feka\ line.

As a consequence, the X-ray observations of these objects would probe
the physical conditions of the central engine with very little
contamination from the radiation originating from the base of the
jet. Alternatively, the difference in flux variability pattern between
the two sources can be explained assuming 
that there are at least two different processes simultaneously at work
in 3C~390.3: one (not related to the jet) causing the smooth secular
decay observed throughout the X-ray monitoring and another (possibly
connected to the jet) giving rise to the small-amplitude flares
superimposed to the long-term variation in the 3C~390.3 light curve
(and to the erratic flux variation displayed by the 3C~120 light
curve). The idea of a contribution from two different physical
processes to the variability of 3C~390.3 is supported by the different
variability behavior in he first (smooth flux decline) and
second part (flux roughly constant) of the light curve, as seen in the
color-color plot (see Figure~\ref{figure:color}).

Other important findings from the temporal analysis are obtained by
applying the fractional variability analysis to energy-selected light
curves and to time-selected intervals of the total light curve. From
the former analysis we found that the fractional variability amplitude
is strongly anticorrelated with the energy band in both sources. 
This behavior
is qualitatively in agreement with the scenario where X-rays are
produced by Compton upscattering in a hot corona located above the
accretion disk, which emits the primary soft flux (e.g., Haardt \&
Maraschi 1991). In this scenario the hard photons are produced by a
larger number of scatterings and therefore their variability is washed
out. The result of the fractional variability analysis of
energy-selected light curves is also important from another point of
view, since it reinforces the claim that the X-rays observed in
3C~390.3 and 3C~120 originate from the central region and not from the
base of a putative X-ray jet. In fact, a similar analysis carried out
on blazars (e.g., Fossati et al. 2000, Chapman et al. 2002), whose
emission is thought to be dominated by non-thermal radiation from
relativistic jets seen at small angles to the line of sight, shows an
opposite trend, with the flux variability increasing with the energy,
and can be simply explained in terms of shorter cooling time for more
energetic electrons.

From the time-resolved fractional variability analysis, 3C~390.3 does
not show any significant correlation, while the fractional variability
of 3C~120 is weakly anticorrelated with the count rate.
While the existence of an
anticorrelation between source luminosity and variability amplitude
among radio-quiet AGN is well known (e.g., Lawrence \& Papadakis 1993,
Nandra et al. 1997, Leighly 1999), to the best of our knowledge
3C~120 represent the first example where this
inverse correlation is found for individual objects. 
This kind of analysis, along with the use of other techniques
not generally applied to AGN light curves (i.e., Gliozzi et al. 2002)
can potentially provide useful information to advance
our understanding of  the X-ray variability in AGN. 
However, the lack of
long X-ray light curves of AGNs with high signal-to-noise ratio
prohibits us from carrying out a systematic analysis on a large
sample. It is worth noting that a similar study was recently carried
out by Turner et al. (2001) on a 35 day ASCA observation of the bright
Narrow-Line Seyfert 1 galaxy Akn 564, and no clear correlation was
found between the fractional variability and X-ray flux for this
source.  On the other hand, in jet dominated AGN $F_{\rm var}$ is
positively correlated with the count rate (e.g., Zhang et al. 1999,
Chapman et al. 2002).

Another remarkable result is the pronounced spectral variability, as
shown by the large-amplitude variations of the X-ray colors, and the
anti-correlation between the color and the local mean count rate. A
similar trend with the spectrum becoming steeper as the flux increases
was already noticed in sub-sample of hard X-ray-selected AGN (Grandi
et al. 1993), and recently confirmed for Seyfert galaxies by Markowitz
\& Edelson (2001) and Papadakis et al. (2002). This trend can be
naturally explained in terms of a variation of the spectral slope,
which is predicted by thermal Comptonization models. However, not all
AGNs show the same spectral behavior (e.g., Grandi et al. 1993) with
some objects showing no energy dependence of their variability (e.g.,
Vaughan 2001) or even a hardening of the spectrum as the source gets
brighter (e.g., Gliozzi et al.  2001).

The main results derived from the time-averaged spectral analysis are
(a) a strong and 
resolved \feka\ line in 3C~390.3 and 3C~120, and (b)
a reflection component that is quite weak (unnecessary in the case of
3C~390.3).  The spectral parameters of 3C~390.3 and 3C~120 are roughly
consistent with previous studies (e.g., Wo\'zniak et al. 1998, Grandi
et al. 1999, Eracleous et al. 2000, Zdziarski \& Grandi 2001), with
the Compton reflection component weaker and the \feka\ line stronger
than the values reported in literature.  A common explanation for the
usually weak reflection component found in radio--loud AGNs is that
the solid angle subtended by the largely neutral reprocessor is
smaller. Since the reprocessor is widely thought to be a standard
optically-thick accretion disk, this has lead to speculation that the
standard disk might be truncated with an optically-thin flow in the
inner region. An alternative interpretation, recently proposed by
Ballantyne et al. (2002), is that the accretion disk is ionized. In
order to test such hypothesis we have fitted the time-averaged spectra
of 3C~390.3 and 3C~120 with the constant-density ionized-disk model of
Ross \& Fabian (1993, Ballantyne et al. 2001) and found a
statistically significant improvement in the spectral fits with
respect to those obtained with the {\tt pexrav} model.  The
time-averaged spectral analysis thus favors the presence of an ionized
disk, but with a moderate ionization parameter ($\xi\simeq 300$). Following
Ballantyne et al. (2002), we also tried to fit the \rxte\ spectra of 
3C~390.3 and 3C~120 keeping the
reflection fraction fixed to 1; however, the resulting $\chi^2$ were not
acceptable in either cases.

The monitoring strategy consisting of short, closely spaced, snapshots
on a long temporal baseline, along with the rather smooth changes in
the X-ray flux shown by 3C~390.3, allowed us to search for
correlations or delays in the variability pattern of the continuum and
the \feka\ line.
The
detection of correlated or delayed line and continuum variability
would be a very powerful diagnostic of the location of the line
emitting region. Unfortunately, though, neither our results nor those
of other studies of BLRGs have shown such an effect. In the case of our study,
the S/N of the \feka\ line in time resolved spectra is not high enough
to show the line variability expected from the observed continuum
variations.  
Other authors (e.g., W\`ozniak et al. 1998 for 3C~390.3, Grandi et
al. 2001 for 3C~382) have claimed that the line intensity in BLRGs
does not track the variations of the continuum, but their conclusions
are based on very sparsely sampled light curves (6 and 3
observations, respectively, spanning a decade) taken with different
instruments. Moreover, the values of the line flux might be affected
by the different spectral models used for the fit of the continuum and
by the different calibrations of the instruments. Thus, the
above conclusions can be stated more precisely by saying that the
light-crossing time of the Fe~K$\alpha$-emitting region is longer than
the duration of a typical observation, which is of the order of a
day. This is not a very strong constraint if one notes that the
light-crossing time of the inner accretion disk ($r\sim 10GM/c^2$) of
$10^9$~M$_{\odot}$ black hole is approximately 0.6~days.

Interesting results are found by plotting spectral parameters
versus the 2--10 keV flux. First, we found that the spectral
variability (already inferred by the color analysis) is due to
intrinsic changes in the photon index, which gets steeper as the flux
increases. A similar spectral behavior was already noticed in 3C~120
by Halpern (1985), who interpreted it in terms of beamed synchrotron
emission from a relativistic jet. But in view of our results and our 
discussion above, emission from the jet makes a very small contribution 
to the observed flux. Therefore, we prefer a scenario in which such 
spectral variability can be simply explained with the standard scenario of
thermal Comptonization from a hot corona, widely accepted for
radio-quiet AGN.  Even more interesting is the \feka\ line flux, which
shows an initial increase as the 2--10 keV flux increases and a
stationary low level at higher values of the X-ray continuum. 
However, the large errors associated with
the time-resolved spectral analysis suggest that such results should
be regarded with caution; yet these results should serve as a
motivation to perform more sensitive X-ray monitoring observations
with new generation X-ray satellites as \xmm.

\section{Conclusion}

A detailed comparison of the temporal and spectral variability
properties of BLRGs with those of jet-dominated and of radio-quiet AGN
leads to two important results: (a) the jet contribution to the X-ray
flux in BLRGs is negligible, and (b) BLRGs qualitatively behave
according to the standard scenario of thermal Comptonization, widely
accepted for radio-quiet AGN. However, the observational data
available at the moment do not give a conclusive answer to the
question of the structure of the inner accretion disks of BLRGs in
particular and radio-loud AGNs in general. On one hand, the X-ray
variability properties of BLRGs seem to be similar to those of Syefert
galaxies suggesting a common emission process. Moreover, model fits to
the \asca\ X-ray spectrum of 3C~120 (Ballantyne et al. 2002) favor a
highly ionized accretion disk. On the other hand, the high-S/N \rxte\
spectra presented here are best fitted with models in which the disk
ionization is modest. In addition, \asca\ observations of BLRGs
such as 3C~390.3 and 3C~445 (e.g., Eracleous, Halpern, \& Livio 1995;
Sambruna et al. 1998) suggest that the \feka\ lines of these objects
come from low-ionization gas. It is unfortunate that the easily
measurable quantities, such as the \feka\ EW and the strength of the
Compton reflection component, are ambiguous indicators of the
ionization state of the disk.

To further address the above issues and to test the hypothesis of the
presence of an ionized accretion disk, a crucial step forward may be
achieved by future X-ray monitoring campaigns and/or high-S/N spectra
with \xmm. In fact the possibility of long uninterrupted observations
combined with the superior sensitivity of EPIC should allow one to perform an
accurate time-resolved analysis. Moreover, the broad-band X-ray
spectrum extending down to 0.2 keV would be important to assess the
presence of a moderately ionized disk, which is supposed to show
strong emission lines (C,N,O) in the soft X-ray range.
 
\acknowledgements 

MG and RMS acknowledge financial support by NASA LTSA grant NAG5-10708.
Support fron NASA grant NAG5-10243 is also acknowledged. ME
acknowledges support from NASA grants NAG5-7733, NAG5-8369, and
NAG5-9982. During the early stages of this project, ME was based at
the University of Califronia, Berkeley and was supported by Hubble
Fellowship grant HF-01068.01-94A, awarded by the Space Telescope
Science Institute.




\begin{deluxetable}{lcrccc}
\tablenum{1}
\tablewidth{6.5in}
\tablecolumns{6}
\tablecaption{Targets: Basic Properties and Observation Log}
\tablehead{
\colhead{} &
\colhead{} &
\colhead{$i$\tablenotemark{a}} &
\colhead{$N_{\rm H,Gal}$} &
\colhead{Start Time (UT)} &
\colhead{End Time (UT)} \\
\colhead{Object} &
\colhead{$z$} &
\colhead{($^{\circ}$)} &
\colhead{$({\rm cm^{-2}})$} &
\colhead{(yy/mm/dd hh:mm)} &
\colhead{(yy/mm/dd hh:mm)} 
}
\startdata
3C~390.3 & 0.056 & $19\leq i\leq 33$ & $3.74\times 10^{20}$ & 1996/05/17 14:14 & 1996/07/12 21:43 \\
3C~120   & 0.033 & $1\leq i\leq 14$  & $1.20\times 10^{21}$ & 1997/01/10 11:43 & 1997/03/08 05:14 \\
\enddata
\tablenotetext{a\;}{The inclination angle of the radio jet (Eracleous \& Halpern 1998)}
\end{deluxetable}


\tablecaption{Observation Details: Total Exposure Times and Count Rates}
\begin{deluxetable}{lcccccc}
\tablenum{2}
\tablewidth{4.7in}
\tablecolumns{6}
\tablehead{
\colhead{} &
\colhead{} &
\multicolumn{2}{c}{PCA (2--20 keV)} &
\colhead{} &
\multicolumn{2}{c}{HEXTE (20--50 keV)} \\
\noalign{\vskip -6 pt}
\colhead{} &
\colhead{} &
\multicolumn{2}{c}{\hrulefill} &
\colhead{} &
\multicolumn{2}{c}{\hrulefill} \\
\colhead{} &
\colhead{} &
\colhead{Exposure} &
\colhead{Count Rate} &
\colhead{} &
\colhead{Exposure} &
\colhead{Count Rate} \\
\colhead{Object} &
\colhead{} &
\colhead{(ks)} &
\colhead{$({\rm s^{-1}})$} &
\colhead{} &
\colhead{(ks)} &
\colhead{$({\rm s^{-1}})$}
}
\startdata
3C~390.3 && 132.14 & $10.53\pm0.02$ && 42.72 & $0.19\pm0.02$ \\
3C~120   && 126.18 & $18.53\pm0.02$ && 40.23 & $0.57\pm0.05$ \\
\enddata
\end{deluxetable}


\tablecaption{Fractional Variability of 3C~390.3 and 3C~120}
\begin{deluxetable}{llclll}
\tablenum{3}
\tablewidth{6in}
\tablecolumns{5.8}
\tablehead{
\colhead{} &
\colhead{Band} &
\colhead{$\langle rate\rangle$} &
\colhead{$F_{\rm var}$} &
\colhead{$\chi^2_{\rm red}$, P$_{\chi^2}$} &
\colhead{$r$, P$_c$} 
\\
\colhead{Object} &
\colhead{(keV)} &
\colhead{$({\rm s^{-1}~PCU^{-1}})$} &
\colhead{(5760 s)}  &
\colhead{}   & 
\colhead{}   
}
\startdata
3C 390.3 & 2--5  & 0.43 &  $(30.8\pm1.9)\times 10^{-2}$ &2.6, 7\% & -0.999, 2\% \\
         & 5--7  & 0.74 &  $(28.3\pm1.6)\times 10^{-2}$ \\
         & 7--11 & 1.06 &  $(25.5\pm1.5)\times 10^{-2}$ \\
\noalign{\hrule} 
3C 120   & 2--5  & 1.09 &  $(8.9\pm0.6)\times 10^{-2}$ &6.4, $\ll1\%$ & -0.975, 10\%\\
         & 5--7  & 1.75 &  $(8.1\pm0.5)\times 10^{-2}$  \\
         & 7--11 & 2.35 &  $(6.4\pm0.4)\times 10^{-2}$  \\
\enddata
\end{deluxetable}


\begin{deluxetable}{llcc}
\tablecaption{Results of Model Fits to Time-Averaged Spectra}
\tablenum{4}
\tablewidth{6in}
\tablecolumns{4}
\tablehead{
\colhead{Model} &
\colhead{Parameters} &
\colhead{3C~390.3} &
\colhead{3C~120} 
}
\startdata
Simple Power-Law         & Photon index, $\Gamma$                    & $1.70\pm 0.01$          & $1.84\pm 0.01$\\
                         & Reduced $\chi^2$/d.o.f                    &  1.06/47                & 1.80/61\\
\noalign{\hrule}
Broken Power-Law         & Low-energy photon index, $\Gamma1$        & $1.70\pm 0.01$          & $1.81\pm 0.01$\\
                         & Break energy (keV)                        & $30^{+80}_{-10}$        & $9\pm 1$\\
                         & High-energy photon index, $\Gamma2$       & $1.40\pm 0.05$          & $1.69^{+0.03}_{-0.04}$\\
                         & Reduced $\chi^2$/d.o.f                    &  0.86/45                & 0.83/59\\
\noalign{\hrule}
Power-Law Plus           & Photon index, $\Gamma$                    & $1.72\pm 0.02$          & $1.83^{+0.05}_{-0.06}$\\
Neutral Disk Reflection  & Folding energy\tablenotemark{a} (keV)     & 14,200 (140)                 & 130  (50)                  \\
                         & Reflector solid angle, $\Omega/2\pi$    & $0.1\pm 0.1$            & $0.4\pm 0.2$    \\
                         & Inclination angle, $i$ (deg)  $^{\rm b}$  & 33                      & 11 \\
                         & Reduced $\chi^2$/d.o.f                    & 1.05/44                 & 0.91/58\\
\noalign{\hrule}
Power-Law Plus           & Photon index, $\Gamma$                    & $1.71\pm 0.01$          & $1.84\pm 0.01$\\
Ionized Disk Reflection  & Log(ionization parameter), $\log\xi$      & $2.5^{+0.2}_{-0.3}$     & $2.4^{+0.1}_{-0.6}$\\
                         & Reflector solid angle, $\Omega/2\pi$    & $0.2^{+0.1}_{-0.2}$     & $0.3^{+0.1}_{-0.2}$ \\
                         & Reduced $\chi^2$/d.o.f                    &  0.94/45                & 0.83/59
\enddata
\tablenotetext{a} {Best fitting values and 90\% confidence lower limits obtained with the {\tt steppar} procedure.}
\tablenotetext{b}{Best fitting value.}
\end{deluxetable}




\tablecaption{Parameters of Gaussian Emission-Line Models}
\begin{deluxetable}{lllllll}
\tablenum{5}
\tablewidth{6.1in}
\tablecolumns{7}
\tablehead{
\colhead{Parameter} &
\multicolumn{2}{c}{3C~390.3} &
\multicolumn{2}{c}{3C~120} \\
\noalign{\vskip -6 pt}
\colhead{} &
\multicolumn{2}{c}{\hrulefill} &
\multicolumn{2}{c}{\hrulefill} \\
\colhead{} &
\colhead{Power-law} &
\colhead{Ionized-disk} &
\colhead{Power-law} &
\colhead{Ionized-disk} \\
}
\startdata
Rest energy (keV) & $6.42^{+0.09}_{-0.07}$ & $6.29^{+0.16}_{-0.31}$ 
&$6.39^{+0.04}_{-0.05}$ &$6.21^{+0.19}_{-0.23}$\\
Rest energy dispersion, $\sigma$ (keV)   & $0.43^{+0.14}_{-0.09}$ & $0.26^{+0.17}_{-0.26}$ &
$0.37^{+0.06}_{-0.04}$ & $0.16^{+0.16}_{-0.16}$\\
Line flux ($10^{-5}~{\rm s^{-1}~cm^{-2}}$) & $5.2^{+0.9}_{-0.8}$  & $2.6^{+2.3}_{-1.1}$ 
& $9.9^{+0.6}_{-0.6}$  & $3.2^{+4.3}_{-1.1}$  \\
Equivalent width (eV)  & $170\pm 30$    & $82^{+73}_{-33}$        
& $150^{+8}_{-10}$ & $45^{+58}_{-16}$      \\
\enddata
\end{deluxetable}


\begin{deluxetable}{cccc}
\tablecaption{Time Intervals Used for Time-Resolved Spectral Analysis}
\tablenum{6}
\tablewidth{5.5in}
\tablecolumns{4}
\tablehead{
\colhead{Start Time (UT)} &
\colhead{End Time(UT)} &
\colhead{Exposure} &
\colhead{2--10 keV Flux} \\
\colhead{(yy/mm/dd hh:mm)} &
\colhead{(yy/mm/dd hh:mm)} &
\colhead{(ks)} &
\colhead{$(10^{-11}{\rm erg~ cm^{-2} ~s^{-1}})$}
}
\startdata
96/05/17 14:14 & 96/05/21 13:33 & 11.84 & 3.70 \\
96/05/22 20:54 & 96/05/26 15:08 & 10.30 & 3.12 \\
96/05/27 16:05 & 96/05/28 21:47 & 12.33 & 3.22 \\
96/06/01 15:50 & 96/06/05 10:12 & 13.22 & 2.90 \\
96/06/06 19:09 & 96/06/10 14:48 & 13.25 & 2.20 \\
96/06/11 12:35 & 96/06/15 03:33 & 14.26 & 1.98 \\
96/06/18 22:13 & 96/06/27 00:21 & 27.28 & 1.74 \\
96/06/27 19:55 & 96/07/02 21:02 & 21.02 & 1.84 \\
96/07/07 13:43 & 96/07/12 22:33 & 8.64  & 2.25 \\
\enddata
\tablenotetext{a}{Calculated assuming a power law plus Gaussian line model.}
\end{deluxetable}


\begin{deluxetable}{cccc}
\tablecaption{Time-resolved spectral parameters}
\tablenum{7}
\tablewidth{4in}
\tablecolumns{4}
\tablehead{
\colhead {} &
\colhead {Energy} &
\colhead {Line} &
\colhead {Equivalent} \\
\colhead {Photon} &
\colhead {Dispersion} &
\colhead {Flux} &
\colhead {Width} \\
\colhead {Index} &
\colhead {(keV)} &
\colhead {$(10^{-5}{\rm cm^{-2} ~s^{-1}})$} &
\colhead {(eV)} 
}
\startdata
$1.79\pm 0.04$         & $0.3_{-0.3}^{+0.6}$ & $7_{-3}^{+4}$  & $140_{-70}^{+90}$   \\
$1.73\pm 0.03$         & $0.3_{-0.3}^{+0.6}$ & $5_{-2}^{+3}$  & $130_{-60}^{+80}$   \\
$1.75\pm 0.03$         & $1.5_{-1.0}^{+1.1}$ & $6_{-2}^{+3}$  & $150_{-50}^{+70}$   \\
$1.71\pm 0.03$         & $0.2_{-0.2}^{+0.4}$ & $6\pm 2$       & $150_{-40}^{+70}$   \\
$1.71\pm 0.04$         & $0.8\pm 0.3$        & $9\pm 3$       & $340_{-110}^{+120}$ \\
$1.62_{-0.05}^{+0.04}$ & $0.7\pm 0.4$        & $7_{-3}^{+4}$  & $270_{-110}^{+150}$ \\
$1.60_{-0.04}^{+0.03}$ & $0.2_{-0.2}^{+0.3}$ & $4_{-1}^{+2}$  & $180_{-50}^{+70}$   \\
$1.69_\pm 0.04$        & $0.4_{-0.4}^{+0.3}$ & $5_{-1}^{+3}$  & $200_{-60}^{+110}$  \\
$1.69_{-0.05}^{+0.11}$ & $1.4_{-1.4}^{+2.1}$ & $7_{-6}^{+28}$ & $200_{-200}^{+900}$ \\
\enddata
\end{deluxetable}


\clearpage



\end{document}